\documentclass[12pt]{article}

\usepackage{booktabs}
\usepackage{color}
\usepackage{latexsym}
\usepackage{amsmath}
\usepackage{array}
\usepackage{amsthm}
\usepackage{amssymb}
\usepackage{amsfonts}
\usepackage{graphicx}
\usepackage{epsfig}
\usepackage{lscape}
\usepackage{natbib}
\usepackage{geometry}
\usepackage{setspace}
\usepackage{multirow}
\usepackage{calc}
\usepackage{ragged2e}
\usepackage[hang,flushmargin]{footmisc}
\usepackage[colorlinks=true, linkcolor=blue, citecolor=blue, urlcolor=blue]{hyperref}
\usepackage{comment}
\usepackage{caption}
\usepackage[utf8]{inputenc}   
\usepackage[T1]{fontenc}      

\usepackage{subfig}

\usepackage{chngcntr}
\counterwithout{table}{section}

\begin{document}

\title{On the Identification of Diagnostic Expectations: Econometric Insights from DSGE Models}
\date{May 18 2026}
\author{
  Jinting Guo\thanks{
    \parbox[t]{0.95\textwidth}{
      PhD candidate at Goethe University Frankfurt. E-mail: \texttt{jinting.guo@stud.uni-frankfurt.de}. \\
I am deeply indebted to Denis Tkachenko for his guidance. I thank my supervisors, Michael Binder and Alexander Meyer-Gohde, for their invaluable supervision. I am also grateful to Uwe Hassler, Ina Krapp, Yulei Luo, Alberto Martin, Hashem Pesaran, Davide Raggi, Yuliya Rychalovska, Ludwig Straub, Penghui Yin, Donghoon Yoo and the members of Monetary Policy and Analysis Division at the Bundesbank for helpful discussion and comments. I further benefited from feedback provided by participants at the 15th RCEA Bayesian Econometrics Workshop, the 13th Annual Conference of the International Association for Applied Econometrics, the 2nd Frankfurt Summer School and the 2025 European Winter Meeting of the Econometric Society. Any remaining errors are my own.
    }
  }
}

\maketitle

\begin{abstract}
This paper shows that diagnostic expectations (DE) and rational expectations (RE) are not observationally equivalent in dynamic stochastic general equilibrium (DSGE) models. Using the frequency-domain framework of \citet{qu2012identification,qu2017global}, I show that no RE parameterization yields the DE-implied autocovariance structure of the macroeconomic observables considered in either small- or medium-scale DSGE models, even after structural frictions and shock processes are reparameterized. Incorporating DE preserves overall identification but weakens the identification of shock variances. In the medium-scale model, among the frictions, wage rigidity emerges as most important for generating the benchmark DE model dynamics.

\bigskip


\end{abstract}

\clearpage
\newpage

\section{Introduction} \label{intro}

Standard dynamic stochastic general equilibrium (DSGE) models have long assumed rational expectations (RE), but diagnostic expectations (DE) have emerged as a plausible alternative. \citet{bordalo2018diagnostic} formalize DE as subjective beliefs under which agents overweight future outcomes that appear more likely in light of incoming information. This mechanism generates extrapolative beliefs and systematic forecast errors. When embedded in DSGE models, DE provide an additional propagation channel for macroeconomic fluctuations \citep{l2024incorporating}. To assess DE as a meaningful alternative to RE, however, an important question is whether DE-DSGE models are empirically distinguishable from otherwise comparable RE models. This paper shows that they are: models under DE generate observable macroeconomic dynamics that cannot be replicated by reparameterizing the corresponding RE models, even when shock processes and nominal, real, and information frictions are allowed to adjust.

To establish this result, this paper applies the frequency-domain identification framework of \citet{qu2012identification,qu2017global} by comparing the joint second-order dynamics implied by the DE model to those implied by its closest RE counterpart. These dynamics are summarized by the spectrum, the frequency-domain representation of the autocovariance structure of the observables. The DE benchmark parameterization is fixed at the posterior mean obtained from Bayesian estimation on quarterly U.S.\ macro data, and the closest RE counterpart is obtained by minimizing the Kullback-Leibler (KL) divergence of the RE-implied spectrum from the DE benchmark spectrum. A strictly positive KL divergence rules out observational equivalence between DSGE models under DE and RE for the observables considered.

The frequency-domain framework is also used to study global identification within the DE model. Whereas the cross-model DE-RE comparison examines whether any RE parameterization can replicate the DE benchmark dynamics, the global identification exercise examines whether any alternative DE parameterization, possibly far from the benchmark, can do so. Without global identification, the model parameters, including the diagnosticity parameter $\theta$, which governs the strength of the belief distortion, could not be uniquely inferred from the observable dynamics.

I first apply both exercises to a small-scale DSGE model, which provides a clear starting point. Using output, inflation, and the nominal interest rate as observables, I show that the diagnosticity parameter is locally and globally identified at the posterior mean. Although introducing DE preserves overall identification, it weakens the identification of shock variances. The intuition is that DE affect the policy functions only through shock impact coefficients while leaving the autoregressive dynamics unchanged. For the cross-model exercise, no RE parameterization in the feasible space is observationally equivalent to the DE benchmark, and the two models are distinguishable at empirically relevant sample sizes. The impulse responses illustrate some aspects of this difference in the time domain: for a positive government spending shock, the DE model generates hump-shaped dynamics in output, inflation, and the nominal interest rate, whereas the closest RE counterpart generates monotonic responses with smaller peak magnitudes. The subjective real interest rate displays an sharper difference. Under DE, it falls to negative on impact and then reverses upward, whereas the closest RE counterpart generates only a positive and monotone response. Intuitively, in line with the mechanism emphasized by \citet{l2024incorporating}, DE agents extrapolate the inflationary effect of the shock into expected future inflation, which lowers the subjective real rate on impact.  As the extrapolative component dissipates and the Taylor-rule response raises the nominal interest rate, the subjective real rate reverses upward.

I then turn to the medium-scale DSGE model of \citet{l2024incorporating}. With rich nominal, real, and information frictions, the RE specification contains propagation channels that can in principle generate amplification and hump-shaped impulse responses similar to those associated with DE. This provides a more demanding test of whether the DE model is empirically distinguishable from the RE counterpart. Nevertheless, DE remain globally identified, and no RE parameterization replicates the spectrum of output, consumption, investment, real wages, labor, inflation, and the nominal interest rate jointly. To minimize the KL divergence, the closest RE model moves toward weaker internal propagation, more flexible price and wage adjustment, more elastic labor supply, a more informative signal, and a more dovish monetary policy rule. These adjustments help the RE model mimic some features of the DE benchmark, but they do not eliminate the divergence.

The medium-scale impulse responses help clarify which features of the DE spectrum the closest RE model cannot replicate. Following a positive government spending shock, the subjective real rate in the closest RE model also exhibits an initial decline followed by a reversal, indicating that with the rich frictions, the medium-scale RE-DSGE model can partially mimic the real rate dynamics under DE. However, the RE response has a smaller impact decline and a weaker subsequent reversal than the DE benchmark. A contractionary monetary policy shock provides a complementary case. Because the shock directly moves the nominal interest rate, the subjective real rate channel plays a more limited role in distinguishing DE from RE. The main discrepancy is instead an impact-persistence trade-off: for output, consumption, investment, and real wages, the closest RE model can match either the impact magnitude or the persistence of the DE model responses to the monetary policy shock, but not both jointly. Across both shocks, the closest RE model also smooths out the kink that DE generate in inflation dynamics.

Finally, having established that models under DE and RE are not observationally equivalent, I identify which structural frictions are empirically most important for generating the DE benchmark dynamics. The main result is that wage rigidity plays a prominent role, and that this role is specific to matching the DE benchmark dynamics rather than a generic feature of this medium-scale model. Moreover, DE and wage rigidity act more as complements than substitutes in generating the benchmark dynamics. Intuitively, sticky real wages slow the adjustment of marginal cost after a demand shock, so the gap between demand and supply closes only gradually, leaving room for diagnostic extrapolation to keep propagating into inflation and real activity beyond the impact period. When wages are flexible, this gap closes quickly and the diagnostic wedge has little to amplify. This finding is consistent with and complementary to \citet{l2024incorporating}, who show in a small-scale setting that DE interact with price rigidity to generate excess output volatility. 

\textbf{Related literature.} \,\,This paper relates to two strands of literature: the literature on DE and the broader research on the identification of structural parameters in structural macroeconomic models. The first strand relates to the growing body of research on DE. \cite{bordalo2018diagnostic} (BGS) introduce DE to explain several features of credit cycles, demonstrating how agents' psychological tendency to overweight representative future outcomes amplifies economic fluctuations. \cite{bordalo2020overreaction} explore DE with dispersed information, finding that agents overreact to private signals while underreacting to consensus forecasts. \cite{bordalo2021diagnostic} incorporate price learning and speculative behavior, accounting for the underreaction-overshooting-crash pattern in asset price bubbles, while \citet{bordalo2021real} demonstrate that embedding DE helps account for financial reversals in business cycle models. \cite{yin2023diagnostic} examine DE under incomplete information and document excess consumption sensitivity consistent with survey evidence. More recently, \cite{bianchi2024diagnostic} show that a DE-based RBC model better replicates boom-bust cycles compared to its rational expectations counterpart. \cite{na2025overreaction} apply this approach to study countercyclical external balances in emerging markets. \citet{l2024incorporating} incorporate DE into a New Keynesian framework, providing a foundation for my identification analysis. While recent contributions propose smooth DE that distortion vary with uncertainty \citep{bianchi2024smooth}, this paper employs the standard BGS framework, which remains the theoretical foundation for most research in this area.

Another strand of research focuses on the identification of DSGE models. \cite{canova2009back} highlight that observational equivalence, partial, and weak identification are widespread in DSGE models, often arising from an ill-behaved mapping between structural parameters and solution coefficients. \cite{iskrev2010local} develops a rank condition providing a sufficient condition for local identification, while \cite{komunjer2011dynamic} propose a Jacobian rank condition offering necessary and sufficient criteria. \cite{qu2012identification} extend this literature by formulating a frequency-domain rank condition for local identification. Additionally, \cite{koop2013identification}  introduce prior-posterior comparison and posterior learning rate indicators as tools for assessing local identification. Research on global identification has emerged more recently. \cite{qu2017global} address this issue in the frequency domain by examining the KL divergence between two DSGE models, providing a framework to assess identification beyond local conditions. More recently, \cite{kocikecki2023solution} proposes an analytical solution for global identification using Gr\"obner basis methods, offering a systematic approach to solving polynomial restrictions in DSGE models.

This paper follows the frequency-domain identification approach of \citet{qu2017global}, which is well suited for comparing the second-order dynamics implied by alternative model structures, such as DSGE models under DE and RE. This is crucial for establishing whether models under DE generate dynamics that cannot be replicated under RE. The method has further advantages. It allows identification analysis at specific frequency ranges, enabling a focus on the business cycle frequencies that are most relevant to DSGE models. It also quantifies identification strength parameter by parameter, indicating which can be estimated reliably from the observable dynamics.

The remainder of the paper proceeds as follows. Section~\ref{sec: framework} introduces diagnostic expectations, the frequency-domain identification framework, and the estimation procedure. Section~\ref{Identification}  studies global identification and observational equivalence in the small-scale DSGE model. Section~\ref{Identification_med} extends the analysis to the medium-scale DSGE model and examines which structural frictions are most important for matching the DE benchmark dynamics. Section~\ref{sec: conclusion} concludes and outlines directions for future research.

\section{Identification Framework} \label{sec: framework}
This section presents the framework used to study the identification
of DE. I first show how the log-linearized DSGE system
under DE can be recast in an RE representation,
following \citet{bordalo2018diagnostic} and
\citet{l2024incorporating}. I then introduce the identification methods of \citet{qu2012identification,qu2017global}.
Finally, I describe the Bayesian estimation procedure, based on Sequential
Monte Carlo (SMC) sampling \citep{herbst2014sequential, cai2021online}, which is used to obtain the benchmark parameter vectors for the identification analysis.
\subsection{Diagnostic Expectations} \label{sec: DE_summary}

DE capture the psychological tendency of agents to overweight future outcomes whose likelihood rises in light of recent news. Following \citet{bordalo2018diagnostic}, consider an exogenous state
variable $\omega_t$ that follows an AR(1) process,
\begin{equation*}
\omega_t = \rho \omega_{t-1} + \varepsilon_t,
\qquad
\varepsilon_t \sim N(0,\sigma_\varepsilon^2),
\end{equation*}
where $\rho \in (0,1]$ is the persistence parameter. A future
state is said to be more representative if it is more likely
under the current realized state $\left\{\omega_t=\hat{\omega_t}\right\}$ than under a reference state based on past
beliefs $ \left\{\omega_t=\rho\hat{\omega}_{t-1}\right\}$. Agents with DE form beliefs by inflating the subjective probability of representative states, with the severity of this distortion governed by the parameter $\theta \geq 0$. When
$\theta=0$, DE coincide with RE; when
$\theta>0$, agents systematically extrapolate recent news.

Under Gaussianity, the distorted diagnostic distribution remains
normal, which implies the RE representation
\begin{equation} \label{equ: DE}
E_t^\theta[\omega_{t+1}]
=
E_t[\omega_{t+1}]
+
\theta\Bigl(E_t[\omega_{t+1}] - E_{t-1}[\omega_{t+1}]\Bigr).
\end{equation}
\citet{l2024incorporating} extend this result to multivariate linear
systems, making it possible to incorporate DE
into log-linearized DSGE models. In particular, the DSGE system under
DE admits an RE representation, which allows the model to be solved and estimated using standard linear methods after augmenting
the state vector. The full construction of the diagnostic distribution, together with the
multivariate RE representation and the solution details for the DSGE system, is reported in Appendix~\ref{app: DE_theory}.

\subsection{Local and Global Identification}\label{sec:identification}
 
To study identification, I begin from the log-linearized DSGE system
in the canonical form of \citet{Sims03}:
\begin{equation} \label{Sims}
    \boldsymbol{\Gamma}_0 \boldsymbol{S}_t
    =
    \boldsymbol{\Gamma}_1 \boldsymbol{S}_{t-1}
    + \boldsymbol{C}
    + \boldsymbol{\Psi} \boldsymbol{z}_t
    + \boldsymbol{\Pi} \boldsymbol{\eta}_t,
\end{equation}
where $\boldsymbol{S}_t$ is the state vector,
$\boldsymbol{z}_t$ is the vector of exogenous shocks, and
$\boldsymbol{\eta}_t$ denotes the vector of expectation errors with
$E_t(\boldsymbol{\eta}_{t+1})=0$. Under DE, the RE representation in equation~\eqref{equ: DE} introduces lagged expectation terms, which require augmenting the state vector to include the relevant one- and two-period-ahead expectations (details in the Online Appendix).
 
Throughout the paper, I restrict attention to parameterizations for which the model admits a unique stable solution. The resulting solution can be written as

\begin{equation} \label{eq:solution}
    \boldsymbol{S}_t = \boldsymbol{\Theta}_1 \boldsymbol{S}_{t-1} + \boldsymbol{\Theta}_{\varepsilon} \boldsymbol{\varepsilon}_t,
\end{equation}
where $\boldsymbol{\Theta}_1$ and $\boldsymbol{\Theta}_{\varepsilon}$ are functions of $\boldsymbol{\Gamma}_0$, $\boldsymbol{\Gamma}_1$, $\boldsymbol{\Psi}$, and $\boldsymbol{\Pi}$, all depending on the parameter vector $\boldsymbol{\gamma}$. Mapping the state vector $\boldsymbol{S}_t$ into the observable vector
$\boldsymbol{Y}_t$ via a selection matrix $\boldsymbol{A}(L)$ yields
\begin{equation}
  \boldsymbol{Y}_t
  =
  \boldsymbol{A}(L)(I-\boldsymbol{\Theta}_1L)^{-1}
  \boldsymbol{\Theta}_\varepsilon
  \boldsymbol{\varepsilon}_t
  \equiv
  \boldsymbol{H}(L,\boldsymbol{\gamma})
  \boldsymbol{\varepsilon}_t.
\end{equation}
Following \citet{qu2012identification,qu2017global}, the spectral
density of $\boldsymbol{Y}_t$ is then given by
\begin{equation}\label{eq:spectrum}
f_{\boldsymbol{\gamma}}(\omega)
  =
  \frac{1}{2\pi}
  \boldsymbol{H}(e^{-i\omega};\boldsymbol{\gamma})
  \boldsymbol{\Sigma}_\varepsilon(\boldsymbol{\gamma})
  \boldsymbol{H}(e^{-i\omega};\boldsymbol{\gamma})',
\end{equation}
where $\boldsymbol{\Sigma}_\varepsilon(\boldsymbol{\gamma})$ denotes the covariance matrix of structural shocks.

In the context of spectral analysis, the parameter vector $\boldsymbol{\gamma}$ is locally identifiable from the
second-order properties of $\{\boldsymbol{Y}_t\}$ at a point $\boldsymbol{\gamma}_0$ if there
exists an open neighborhood of $\boldsymbol{\gamma}_0$ such that
\[
  f_{\boldsymbol{\gamma}_0}(\omega)=f_{\boldsymbol{\gamma}_1}(\omega),
  \quad \forall\,\omega\in[-\pi,\pi]
  \qquad \Longleftrightarrow \qquad
  \boldsymbol{\gamma}_0=\boldsymbol{\gamma}_1.
\]
By Theorem~1 of \citet{qu2012identification}, a necessary and
sufficient condition for local second-order identification at
$\boldsymbol{\gamma}_0$ is that
\begin{equation}\label{eq:G_matrix}
  \boldsymbol{G}(\boldsymbol{\gamma}_0)
  =
  \int_{-\pi}^{\pi}
  \left(
  \frac{\partial\,\mathrm{vec}\,f_{\boldsymbol{\gamma}_0}(\omega)}
       {\partial\boldsymbol{\gamma}'}
  \right)^{\prime}
  \left(
  \frac{\partial\,\mathrm{vec}\,f_{\boldsymbol{\gamma}_0}(\omega)}
       {\partial\boldsymbol{\gamma}'}
  \right)
  d\omega
\end{equation}
has full rank, where $\mathrm{vec}$ denotes the column-stacking
operator. 
 
Local identification rules out observational equivalence only in a
neighborhood of $\boldsymbol{\gamma}_0$. Global identification is stronger: it
requires that no other parameter vector in the feasible parameter
space $\Theta$ generates the same spectrum. By Theorem~2 of
\citet{qu2017global}, under regularity conditions,\footnote{See
Assumptions 1, 2, and 4 in \citet{qu2017global}.} global
identification is equivalent to strict positivity of the
KL divergence between the spectra implied by
$\boldsymbol{\gamma}_0$ and any $\boldsymbol{\gamma}_1\neq\boldsymbol{\gamma}_0$. The KL divergence is defined as
\begin{equation}\label{eq:KL}
  KL(\boldsymbol{\gamma}_0,\boldsymbol{\gamma}_1)
  =
  \frac{1}{4\pi}
  \int_{-\pi}^{\pi}
  \left\{
    \mathrm{tr}\bigl(f_{\boldsymbol{\gamma}_1}^{-1}(\omega)f_{\boldsymbol{\gamma}_0}(\omega)\bigr)
    -\log\det\bigl(f_{\boldsymbol{\gamma}_1}^{-1}(\omega)f_{\boldsymbol{\gamma}_0}(\omega)\bigr)
    -n_Y
  \right\}
  d\omega,
\end{equation}
where $n_Y$ is the dimension of $\boldsymbol{Y}_t$. If the model is locally
identified at $\boldsymbol{\gamma}_0$, the global identification condition reduces
to
\begin{equation}\label{eq:KL_global}
  \inf_{\boldsymbol{\gamma}_1\in\Theta\setminus B(\boldsymbol{\gamma}_0)}
  KL(\boldsymbol{\gamma}_0,\boldsymbol{\gamma}_1) > 0,
\end{equation}
where $B(\boldsymbol{\gamma}_0)$ is an open neighborhood of $\boldsymbol{\gamma}_0$.
 
The same framework also applies to identification across model
structures. When the benchmark DE model implies spectrum
$f_{\boldsymbol{\gamma}_0}(\omega)$ and the alternative RE model implies spectrum $f_{\boldsymbol{\phi}}(\omega)$, the two structures generate
observationally distinct dynamics if and only if
\[
\inf_{\boldsymbol{\phi}\in\Phi} KL_{ff}(\boldsymbol{\gamma}_0,\boldsymbol{\phi}) > 0,
\]
as shown in Corollary~3 of \citet{qu2017global}. This cross-model exercise is central to the analysis below, as it allows me to assess whether any RE parameterization within the feasible parameter space can be observationally equivalent to the DE benchmark.
 
Because the KL divergence is computed numerically\footnote{Numerical error arises from three sources. First, the DSGE solution is computed using Sims's \texttt{gensys.m}, whose numerical error is of order $10^{-15}$ \citep{anderson2008solving}. Second, the integral used to construct the KL divergence is approximated numerically. Third, the minimization of the KL divergence is carried out up to a finite tolerance level.}, values below
$10^{-10}$ are treated as zero, following \citet{qu2017global}. To
assess the performance in finite samples, I also report the empirical KL divergence
proposed by \citet{qu2017global}. This measure can be interpreted as the highest power of a test of $f_{\boldsymbol{\gamma}_0}(\omega)$ against an alternative spectrum $f_{\boldsymbol{\phi}}(\omega)$ in finite samples under Gaussianity, with higher values indicating greater divergence between the two spectra. The null hypothesis is that $f_{\boldsymbol{\gamma}_0}(\omega)$ is the true spectrum, while the alternative is given by the spectrum that minimizes the KL divergence, either from an alternative parameterization within the same model or from an alternative model structure. A higher value of KL divergence means that it is easier to distinguish between the two spectra.
 
\subsection{Estimation}\label{sec:estimation}
In this paper, I estimate the small-scale DSGE models using SMC \citep{herbst2014sequential}, which provides the benchmark parameter vector $\boldsymbol{\gamma}_0$ for the identification analysis. SMC gradually transforms a set of weighted particles from the prior distribution into the posterior distribution by passing through a sequence of bridge distributions:
\begin{equation}
\pi_n(\boldsymbol{\gamma})
=
\frac{\left[p(\boldsymbol{Y}_{1:T}\mid \boldsymbol{\gamma})\right]^{\phi_n}p(\boldsymbol{\gamma})}
{\int \left[p(\boldsymbol{Y}_{1:T}\mid \boldsymbol{\gamma})\right]^{\phi_n}p(\boldsymbol{\gamma})\,d\boldsymbol{\gamma}},
\qquad n=1,\ldots,N_\phi,
\end{equation}
where $\phi_n \in [0,1]$ is the tempering parameter, with $\phi_1 = 0$ corresponding to the prior and $\phi_{N_\phi} = 1$ corresponding to the posterior. The log likelihood $\log p(\boldsymbol{Y}_{1:T}\mid\boldsymbol{\gamma})$ is evaluated via the Kalman filter applied to the model's state-space representation. As shown by \cite{herbst2014sequential}, SMC is more efficient than the Random Walk Metropolis-Hastings algorithm and performs well with multimodal posteriors, making it suitable for DSGE estimation where posterior landscapes can be irregular.

\section{Identification analysis for a small-scale DSGE} \label{Identification}
In this section, I study the local and global identification of the diagnosticity parameter $\theta$ in the small-scale DSGE model of
\citet{l2024incorporating}. I also assess whether DE and RE are observationally equivalent in this model. The model is specified as follows:
  \begin{align}
         \hat{y}_t&=E_t^\theta [\hat{y}_{t+1}]-(\hat{i}_t-E_t^\theta[\hat{\pi}_{t+1}])+\theta(\hat{\pi}_t-E_{t-1}[\hat{\pi}_t]) +\hat{g}_t-E_t^\theta[\hat{g}_{t+1}]\\
         \hat{\pi}_t&=\beta E_t^\theta[\hat{\pi}_{t+1}]+\kappa(\hat{y}_t-\hat{a}_t)-\frac{\kappa}{1+\nu} \hat{g}_t\\
\hat{i}_t&=\phi_\pi\hat{\pi}_t+\phi_y(\hat{y}_t-\hat a_t)+\varepsilon_{m,t},
        \end{align}
   with shock processes:
        \begin{align}
          \hat{a}_t&=\rho_a\hat{a}_{t-1}+\varepsilon_{a,t}\\
         \hat{g}_t&=\rho_g\hat{g}_{t-1}+\varepsilon_{g,t}
        \end{align}
Relative to \citet{l2024incorporating}, the specification here
includes an additional monetary policy shock, $\varepsilon_{m,t}$, so
as to square the system.

\subsection{Local and global identification}
In this subsection, I examine the identification properties of the parameter vector at the posterior mean obtained from Bayesian estimation. I estimate the small-scale model using quarterly U.S.\
data from \citet{smets2007shocks}, with three demeaned
observables: output growth, inflation, and the nominal interest rate, so that $(\hat{y}_t-\hat{y}_{t-1},\,\hat{\pi}_t,\,\hat{i}_t)$ are centered around the steady state $(0,0,0)$. Following \citet{qu2017global}, I split the sample into the pre-Volcker period (1960Q1--1979Q2) and the post-1982 period (1982Q4--1997Q4). Because I focus exclusively on the determinate case, I use the latter subsample.
\begin{table}[h]
\centering
\caption{Prior and Posterior Distributions}
\label{tab:prior_posterior}
\vspace*{0.2cm}
\renewcommand{\arraystretch}{1.1} 
\setlength{\tabcolsep}{4pt} 
\begin{tabular}{l l l c c c c c c}
\toprule
\multirow{2}{*}{\textbf{Param.}} 
& \multicolumn{4}{c}{\textbf{Prior}} 
& \multicolumn{2}{c}{\textbf{Posterior DE}} 
& \multicolumn{2}{c}{\textbf{Posterior RE}} \\
\cmidrule(lr){2-5} \cmidrule(lr){6-7} \cmidrule(lr){8-9}
& Description & Dist. & Mean & Std.& Mean & \text{90\% HPD} & Mean & \text{90\% HPD} \\
\midrule
$\theta$       & diagnosticity   & Normal     & 1.00  & 0.30  & 0.57 & [0.41, 0.74] & 0    & --- \\
$\phi_y$       & m.p. rule       & Normal     & 0.50  & 0.25  & 0.11 & [0.07, 0.14] & 0.08 & [0.05, 0.11] \\
$\phi_{\pi}$   & m.p. rule       & Normal     & 1.50  & 0.25  & 1.15 & [0.99, 1.31] & 1.21 & [0.99, 1.38] \\
$\kappa$       & P.C. slope      & Gamma      & 0.05  & 0.025 & 0.12 & [0.08, 0.16] & 0.15 & [0.10, 0.20] \\
$\rho_a$       & persis. tech.   & Beta       & 0.50  & 0.20  & 0.77 & [0.65, 0.91] & 0.56 & [0.41, 0.74] \\
$\rho_g$       & persis. fisc.   & Beta       & 0.50  & 0.20  & 0.93 & [0.91, 0.96] & 0.95 & [0.94, 0.97] \\
$\sigma_a$     & s.d. tech.      & Inv. Gamma & 0.50  & 1.00  & 0.61 & [0.38, 0.83] & 0.91 & [0.60, 1.23] \\
$\sigma_g$     & s.d. fisc.      & Inv. Gamma & 0.50  & 1.00  & 1.79 & [1.42, 2.15] & 1.54 & [1.31, 1.80] \\
$\sigma_m$     & s.d. mon.       & Inv. Gamma & 0.50  & 1.00  & 0.38 & [0.33, 0.44] & 0.39 & [0.33, 0.46] \\
\bottomrule
\end{tabular}

\vspace*{0.3cm}
\begin{minipage}{\textwidth}
\justify
{\small Note: The results were estimated using Dynare version 6.2, with the number of particles (\( N \) in \cite{herbst2014sequential}) set to 3,000 and the number of stages (\( N_{\phi} \) in \cite{herbst2014sequential}) set to 200. The parameter \( \lambda \) is set to 2, with an initial scaling parameter of 0.5 and an initial acceptance rate of 0.25. These settings align with those used by \cite{cai2021online} in the Bayesian SMC estimation for \cite{an2007bayesian}. Since \cite{cai2021online} demonstrated that the gain from increasing the mutation block is limited, I set it to 1.}
\end{minipage}
\end{table}

The prior distributions selected are summarized in Table \ref{tab:prior_posterior}. The prior means for $\theta$, $\phi_y$, $\phi_\pi$, and $\kappa$ align with the calibrations used by \cite{l2024incorporating}. Additionally, following \cite{l2024incorporating}, I calibrate the discount factor $\beta = 0.99$ and the inverse Frisch elasticity $\nu = 2$\footnote{I calibrate \(\beta\) because the data have been demeaned. Since the model is locally identified only when \(\nu\) is fixed, even under RE, I set \(\nu = 2\). My primary objective is to test whether DE can be identified, rather than to assess the model’s overall identification or fit to the data.}. Due to the multimodal nature of the Phillips curve (PC) slope parameter, the prior for $\kappa$, which is inversely related to price rigidity, plays a significant role in its estimation. According to \cite{del2008forming}, two competing views exist regarding price rigidity: one favoring high rigidity and the other low rigidity. Evaluating these perspectives is beyond the scope of this paper; therefore, I adopt a high price rigidity prior consistent with the calibration in \cite{l2024incorporating}, which is supported by extensive literature (e.g., \citealt{schorfheide2008dsge}; \citealt{nakamura2014fiscal}; \citealt{gali2015monetary}; \citealt{jones2021priors}; \citealt{hazell2022slope}).
The shock-related priors used are standard. Table \ref{tab:prior_posterior} reports the posterior mean estimates along with their 90\% HPD intervals. For the subsequent identification analysis, I use the posterior means of both DE and RE models as the respective $\gamma_0$ values.

With the posterior means in hand, I now assess the identification
properties of the model. The $G$-matrix evaluated at the posterior
mean $\gamma_0^{DE}$ has full rank\footnote{The corresponding $G$-matrix under RE also has full rank at
$\gamma_0^{RE}$.}, with its smallest eigenvalue well
above the numerical tolerance of $10^{-10}$. This establishes local identification of the small-scale DSGE model under DE at the posterior mean.

\begin{table}[h!]
\centering
\caption{Parameter values minimizing the KL criterion, HSY (2024) model under DE}
\label{tab: optimizer_DE}
\vspace*{0.2cm}
\begin{tabular}{c c c c c c c c c c c c}
\toprule
 & \multicolumn{4}{c}{\textbf{(a) All parameters can vary}} & \multicolumn{4}{c}{\textbf{(b) $\sigma_a$ fixed}} & \multicolumn{3}{c}{\textbf{(c) $\sigma_a$ and $\sigma_g$ fixed}} \\
\cmidrule(lr){2-5} \cmidrule(lr){6-9} \cmidrule(lr){10-12}
 & $\gamma_0^{DE}$ & $c=0.1$ & $c=0.5$ & $c=1$ & $c=0.1$ & $c=0.5$ & $c=1$ & & $c=0.1$ & $c=0.5$ & $c=1$ \\
\midrule
$\theta$ & 0.57 & 0.63 & 0.81 & 0.38 & 0.61 & 0.38 & 0.21 & & 0.55 & 0.53 & 0.51 \\
$\phi_y$ & 0.11 & 0.11 & 0.12 & 0.13 & 0.11 & 0.14 & 0.09 & & 0.10 & 0.06 & 0.01 \\
$\phi_{\pi}$ & 1.15 & 1.07 & 0.73 & 0.38 & 1.13 & 1.11 & 1.07 & & \textbf{1.25} & \textbf{1.65} & \textbf{2.15} \\
$\beta$ & 0.990 & 0.902 & 0.686 & 0.620 & 0.958 & 0.966 & 0.999 & & 0.999 & 0.999 & 0.999 \\
$\kappa$ & 0.12 & 0.11 & 0.09 & 0.07 & 0.12 & 0.13 & 0.14 & & 0.14 & 0.23 & 0.34 \\
$\rho_a$ & 0.77 & 0.76 & 0.76 & 0.72 & 0.77 & 0.83 & 0.75 & & 0.74 & 0.67 & 0.60 \\
$\rho_g$ & 0.93 & 0.93 & 0.91 & 0.91 & 0.93 & 0.89 & 0.95 & & 0.93 & 0.93 & 0.93 \\
$\sigma_a$ & 0.61 & \textbf{0.71} & \textbf{1.11} & \textbf{1.61} & 0.61 & 0.61 & 0.61 & & 0.61 & 0.61 & 0.61 \\
$\sigma_g$ & 1.79 & 1.71 & 1.54 & 1.41 & \textbf{1.69} & \textbf{1.29} & \textbf{2.79} & & 1.79 & 1.79 & 1.79 \\
$\sigma_m$ & 0.38 & 0.38 & 0.37 & 0.37 & 0.38 & 0.39 & 0.37 & & 0.38 & 0.39 & 0.44 \\
\bottomrule
\end{tabular}
\par
\vspace*{0.3cm}
\begin{minipage}{1.02\textwidth}
\justify
{\small Note: KL denotes $KL_{ff}(\gamma_0^{DE}, \gamma_c)$ with $\gamma_0^{DE}$  corresponding to the benchmark specification. The values are rounded to the second decimal place except for $\beta$. The bold value signifies the binding constraint. }
\end{minipage}
\end{table}

\begin{table}[h!]
\centering
\caption{KL and empirical distances between $\gamma_c$ and $\gamma_0$, HSY (2024) model}
\label{tab: KL_DE}
\vspace*{0.2cm}
\setlength{\tabcolsep}{3pt} 
\begin{tabular}{c c c c c c c c c c c c}
\toprule
 & \multicolumn{4}{c}{\textbf{(a)\small All parameters can vary}} & \multicolumn{3}{c}{\textbf{(b) $\sigma_a$ fixed}} & \multicolumn{3}{c}{\textbf{(c) $\sigma_a$ and $\sigma_g$ fixed}} \\
\cmidrule(lr){2-5} \cmidrule(lr){6-8} \cmidrule(lr){10-12}
 & $c=0.1$ & $c=0.5$ & $c=1$ & & $c=0.1$ & $c=0.5$ & $c=1$ & & $c=0.1$ & $c=0.5$ & $c=1$  \\
\midrule
KL & 2.15E-04 & 5.16E-03 & 1.23E-02 & & 8.05E-04 & 0.0266 & 0.0811 & & 1.50E-03 & 0.0378 & 0.1230 \\
$T=80$ & 0.0565 & 0.1540 & 0.3232 & & 0.0823 & 0.5232 & 0.9727 & & 0.1077 & 0.6653 & 0.9794 \\
$T=150$ & 0.0649 & 0.2515 & 0.5267 & & 0.1048 & 0.7857 & 0.9983 & & 0.1457 & 0.9154 & 0.9999 \\
$T=200$ & 0.0701 & 0.3196 & 0.6438 & & 0.1196 & 0.8873 & 0.9998 & & 0.1712 & 0.9725 & 1.0000 \\
$T=1000$ & 0.1346 & 0.9167 & 0.9992 & & 0.3222 & 1.0000 & 1.0000 & & 0.5092 & 1.0000 & 1.0000 \\
\bottomrule
\end{tabular}
\par
\vspace*{0.3cm}
\begin{minipage}{1.02\textwidth}
\justify
{\small Note: KL denotes $KL_{ff}(\gamma_0^{DE}, \gamma_c)$ with $\gamma_0^{DE}$  given in the column 1 of Table~\ref{tab: optimizer_DE} (a). The empirical distance measure equals $p_{ff}(\gamma_0^{DE}, \gamma_c, 0.05, T)$, where $T$ is specified in the last four rows of the table. }
\end{minipage}
\end{table}
I next conduct the global identification exercise by minimizing $KL(\gamma_0^{DE}, \gamma^{DE})$ under the constraint
$\|\gamma^{DE} - \gamma_0^{DE}\|_{\infty} \geq c$ for the benchmark DE model. The results, shown in
Tables~\ref{tab: optimizer_DE} and~\ref{tab: KL_DE}, indicate that the empirical KL distances consistently exceed 0.05 and increase with
sample size $T$ when all parameters vary, implying no observational equivalence. The corresponding RE results are reported in Appendix~\ref{sec: App_B.2}.\footnote{While direct comparison of absolute KL values between the RE and DE models is not meaningful due to the additional parameter $\theta$, examining the relative identification strength of individual parameters across the two specifications is informative.}

The relative strength of identification is revealing. Under DE, the shock variance parameters $\sigma_a$ and $\sigma_g$ exhibit the weakest identification strength, followed by $\phi_\pi$, whereas under RE the weakest identified parameters are $\beta$ and $\phi_\pi$. This difference has a structural explanation. As shown by the analytical solution in Appendix~\ref{sec: App_B.1} and in the Online Appendix of \citet{l2024incorporating}, $\theta$ enters the policy functions solely through the shock impact coefficients, leaving the autoregressive dynamics unchanged. Introducing $\theta$ therefore
 reduces the sensitivity of the model-implied spectrum to changes in shock variances. This translates into weaker identification for $\sigma_a$ and $\sigma_g$ under DE. The model nonetheless remains globally identifiable, with $\theta$ more strongly identified than the shock variances.

\subsection{Detecting observational equivalence between DE and RE}\label{sec: DE_RE_small}
In this subsection, I assess whether the small-scale model under DE and RE can generate observationally equivalent dynamics for output, inflation, and the nominal interest rate. Following \citet{qu2017global,qu2023using}, I fix the DE model at its posterior mean and search over the feasible parameter space of the RE model that minimizes the KL divergence, imposing $\theta = 0$ while allowing all other parameters to vary freely for the RE model. According to \citet{qu2017global}, DE and RE parameterizations are observationally equivalent only if this minimum KL divergence is zero; a strictly positive value implies that no feasible RE parameterization can replicate the observable spectrum implied by the DE benchmark parameterization. To the best of my knowledge, the framework of \citet{qu2017global} remains the only formal approach for studying identification across distinct model structures, making it well suited to detecting whether any RE parameterization can yield the DE-implied spectrum.

\begin{table}[h!]
\centering
\caption{The closest RE model to the small-scale benchmark DE model}
\label{tab: unconstrained}
\vspace*{0.2cm}
\setlength{\tabcolsep}{10pt} 
\begin{tabular}{p{3cm} p{3cm} p{3cm}}
\toprule
 & \textbf{DE} & \textbf{RE} \\
\midrule
$\theta$ & 0.57 & 0 \\
$\phi_y$ & 0.11 & 0.09 \\
$\phi_{\pi}$ & 1.15 & 1.19 \\
$\beta$ & 0.99 & 0.999 \\
$\kappa$ & 0.12 & 0.16 \\
$\rho_a$ & 0.77 & 0.66 \\
$\rho_g$ & 0.93 & 0.94 \\
$\sigma_a$ & 0.61 & 0.79 \\
$\sigma_g$ & 1.79 & 1.41 \\
$\sigma_m$ & 0.38 & 0.37 \\
\bottomrule
\end{tabular}
\par
\vspace*{0.3cm}
\begin{minipage}{0.75\textwidth}
\justify
{\small Note: Column DE shows posterior means of parameters from the benchmark HSY (2024) model. Column RE shows parameters that minimize $KL_{ff}(\gamma_0^{DE}, \gamma^{RE})$, where $f$ is the spectrum and $\gamma^{RE}$ is the parameter vector under RE.} 
\end{minipage}
\end{table}

\begin{table}[h!]
\centering
\caption{KL and empirical distances between the benchmark and its closest RE counterpart}
\label{tab: KL_unconstrained}
\vspace*{0.2cm}
\setlength{\tabcolsep}{10pt} 
\begin{tabular}{p{5cm} p{5cm}}
\toprule
 & \textbf{Value} \\
\midrule
KL & 0.0711 \\
$T=80$ & 0.9531 \\
$T=150$ & 0.9961 \\
$T=200$ & 0.9994 \\
$T=1000$ & 1.0000 \\
\bottomrule
\end{tabular}
\par
\vspace*{0.3cm}
\begin{minipage}{0.85\textwidth}
\justify
{\small Note: KL divergence and the empirical distance measure are defined as $KL_{ff}(\gamma_0^{DE}, \gamma^{RE})$ and $p_{ff}(\gamma_0, \zeta, 0.05, T)$, where $f$ and $\gamma^{RE}$ are the spectrum and structural parameter vector of the alternative model respectively and $T$ specified in the last four rows of the table. }
\end{minipage}
\end{table}

The results are presented in Tables \ref{tab: unconstrained} and \ref{tab: KL_unconstrained}. The search for minimizers is conducted over a relatively large parameter space: $\gamma^{RE} = [\phi_y,\, \phi_\pi,\, \beta,\, \kappa, \rho_a,\, \rho_g,\sigma_a,\, \sigma_g,\, \sigma_m] \in [(0.01,0.99); (0.1,5);\\ (0.1,0.999); (0.01,3); (0.1,0.99); (0.1,0.99);(0.1,3); (0.1,3); (0.1,3)]$. Despite the generous bou\\-nds on the parameter space, no set of RE parameters matches the spectrum implied by the DE benchmark.
The theoretical KL divergence is 0.0711, and the empirical distance is larger than 0.9 even for $T=80$. DE play a crucial role in generating the observable macroeconomic dynamics implied by the benchmark model, which cannot be replicated under RE.

To compensate for the absence of DE, the RE minimizer steepens the Phillips curve, with $\kappa$ increasing from 0.12 to 0.16. Beyond this, the RE model adjusts mainly through the exogenous shock processes: TFP shocks become more volatile but less persistent, with $\sigma_a$ rising from 0.61 to 0.79 and $\rho_a$ falling from 0.77 to 0.66, while government spending shocks become slightly more persistent but less volatile, with $\rho_g$ increasing from 0.93 to 0.94 and $\sigma_g$ declining from 1.79 to 1.41.
\begin{figure}[h]
    \centering
    \caption{Impulse responses to a government spending shock: DE baseline vs.\ RE (small-scale model)}
    \includegraphics[width=1.05\textwidth, trim=15 10 15 10, clip]{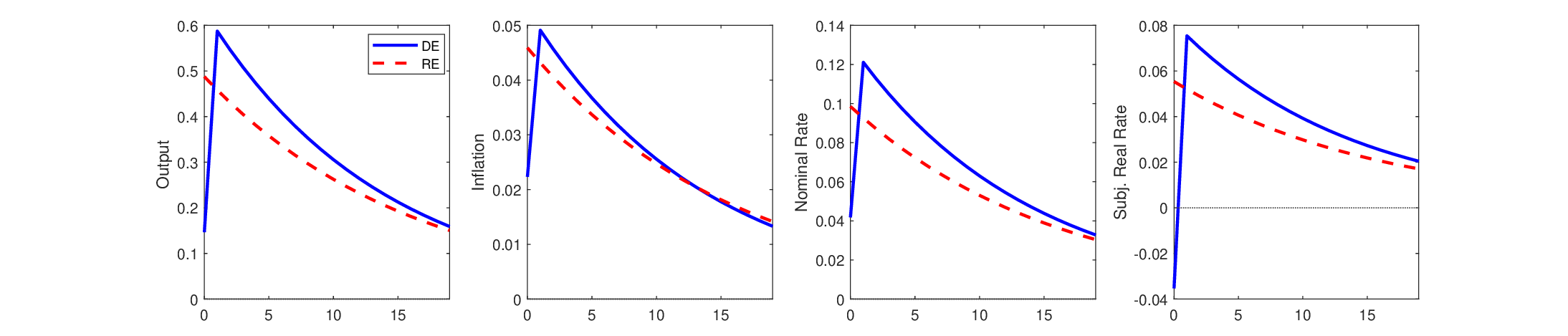}
    \label{fig:IRF_DE_RE}
   \caption*{\small Note: This figure displays impulse responses of output, inflation, the nominal interest rate, and the subjective real interest rate to a one standard deviation positive government spending shock in the small-scale DSGE model. The blue solid lines correspond to the DE model evaluated at the posterior means. The red dashed lines correspond to the RE model with parameters chosen to minimize the KL divergence relative to the DE model (see Table~\ref{tab: unconstrained}). The horizontal axis is in quarters.}
\end{figure}

Figure~\ref{fig:IRF_DE_RE} plots impulse responses for the DE model and its closest RE counterpart, with parameter values reported in Table~\ref{tab: unconstrained}. Following a positive government spending shock, the DE model generates hump-shaped responses in output, inflation, and the nominal interest rate, peaking one quarter after impact before gradually reverting. By contrast, the closest RE counterpart generates monotone responses with smaller peak magnitudes. The subjective real interest rate displays the sharpest difference: under DE, it falls below zero on impact and then reverses upward before decaying, a non-monotone pattern that the closest RE counterpart cannot reproduce.

Intuitively, the mechanism follows the diagnostic Fisher equation emphasized by \citet{l2024incorporating}:
\[
\hat r_t
=
\hat i_t
-
E_t[\hat\pi_{t+1}]
-
\theta\left(E_t[\hat\pi_{t+1}]-E_{t-1}[\hat\pi_{t+1}]\right)
-
\theta\left(\hat\pi_t-E_{t-1}[\hat\pi_t]\right).
\]
A positive government spending shock raises inflation, and diagnostic agents extrapolate this inflationary effect into expected future inflation. This lowers the subjective real rate on impact. The subsequent reversal occurs as the initial diagnostic extrapolative component fades, while the Taylor-rule response raises the nominal interest rate in response to inflationary pressure. At the estimated values, this policy response is strong enough to push the subjective real rate back above zero after its initial decline. Under RE, these diagnostic terms are absent, and the subjective real rate remains positive throughout and decays smoothly from impact.

\section{Identification analysis for a medium-scale DSGE} \label{Identification_med}
I now turn to the medium-scale DSGE model of \citet{l2024incorporating}, which extends the small-scale framework with a richer set of nominal, real, and information frictions. DE remains globally identified at the benchmark posterior mean, and no RE model parameterization replicates the DE model dynamics across seven observables, even with all other structural parameters free to adjust. Among the model's frictions, wage rigidity emerges as by far the most important for matching the DE benchmark spectrum.

\subsection{Global identification}\label{Identification_med_FD}
Specifically, \citet{l2024incorporating} develop a medium-scale DSGE model featuring investment adjustment costs, variable capital utilization, habit formation in consumption, price and wage stickiness, and a noisy signal about permanent productivity. I follow their framework closely, with one modification to the information structure. Following \citet{blanchard2013news}, productivity is decomposed into a nonstationary permanent component and a stationary transitory component, and agents receive a noisy signal about the permanent component rather than observing either component directly. In my implementation, I express the permanent component in growth rates and work with a stationary system; agents observe a noisy signal about permanent productivity growth rather than its level. This reformulation is adopted for numerical tractability. It does not change the economic role of the information friction.
\begin{table}[!htbp]
\centering
\caption{Parameter values minimizing the KL criterion, HSY (2024) model}
\label{tab:kl_params}
\vspace*{0.2cm}
\renewcommand{\arraystretch}{1.05}
\setlength{\tabcolsep}{4pt}
\begin{tabular}{@{}ll c ccc  ccc@{}}
\toprule
& & & \multicolumn{3}{c}{(a) All parameters can vary} 
  & \multicolumn{3}{c}{(b) $\sigma_\mu$ fixed} \\
\cmidrule(lr){4-6} \cmidrule(l){7-9}
Parameter & Description & $\gamma_0^{med}$ 
  & $\,\,\,\,\,c=0.1$ & $\,\,\,\,c=0.5$ & $\,\,c=1.0$ 
  & $c=0.1$ & $c=0.5$ & $c=1.0$ \\
\midrule
$\theta$  & diagnosticity   & 0.72  & 0.72  & 0.72  & 0.73  & 0.72  & 0.73  & 0.74 \\
$\alpha$  & cap.\ share     & 0.13  & 0.13  & 0.13  & 0.13  & 0.13  & 0.13  & 0.13 \\
$h$       & habits          & 0.72  & 0.72  & 0.72  & 0.72  & 0.72  & 0.72  & 0.72 \\
$\frac{\chi''(1)}{\chi'(1)}$ & cap.\ util.\ costs & 5.09  & 5.09  & 5.11  & 5.13  & 5.09  & 5.09  & 5.09 \\[2pt]
$\kappa_p$ & price PC slope & 0.04  & 0.04  & 0.04  & 0.04  & 0.04  & 0.04  & 0.04 \\
$\kappa_w$ & wage PC slope  & 0.01  & 0.01  & 0.01  & 0.01  & 0.01  & 0.01  & 0.01 \\
$\nu$     & inv.\ Frisch elas.\ & 3.71  & 3.70  & 3.68  & 3.65  & \textbf{3.81}  & \textbf{4.21}  & \textbf{4.71} \\
$S''(1)$  & inv.\ adj.\ costs   & 6.93  & 6.95  & 7.02  & 7.11  & 6.92  & 6.89  & 6.86 \\
$\rho_R$  & m.p.\ rule      & 0.58  & 0.58  & 0.58  & 0.58  & 0.58  & 0.58  & 0.59 \\
$\phi_{\pi}$ & m.p.\ rule   & 1.54  & 1.54  & 1.54  & 1.53  & 1.54  & 1.55  & 1.55 \\
$\phi_x$  & m.p.\ rule      & 0.006 & 0.006 & 0.006 & 0.006 & 0.006 & 0.006 & 0.007 \\
\addlinespace[4pt]
\multicolumn{9}{@{}l}{\textit{Autoregressive parameters}} \\[2pt]
$\rho$       & tech.        & 0.85  & 0.85  & 0.85  & 0.85  & 0.85  & 0.85  & 0.85 \\
$\rho_\mu$   & invest.      & 0.31  & 0.31  & 0.31  & 0.30  & 0.31  & 0.31  & 0.31 \\
$\rho_p$     & price m-up   & 0.88  & 0.88  & 0.88  & 0.88  & 0.88  & 0.88  & 0.88 \\
$\phi_p$     & price ma     & 0.58  & 0.58  & 0.58  & 0.58  & 0.58  & 0.58  & 0.57 \\
$\rho_w$     & wage m-up    & 0.99  & 0.99  & 0.99  & 0.99  & 0.99  & 0.99  & 0.99 \\
$\phi_w$     & wage ma      & 0.54  & 0.54  & 0.53  & 0.53  & 0.54  & 0.56  & 0.57 \\
$\rho_{mp}$  & policy       & 0.03  & 0.03  & 0.03  & 0.03  & 0.03  & 0.03  & 0.03 \\
$\rho_g$     & govt.        & 0.94  & 0.94  & 0.94  & 0.94  & 0.94  & 0.94  & 0.94 \\
\addlinespace[4pt]
\multicolumn{9}{@{}l}{\textit{Standard deviations}} \\[2pt]
$\sigma_a$    & tech.       & 1.43  & 1.43  & 1.43  & 1.44  & 1.43  & 1.43  & 1.44 \\
$\sigma_s$    & tech.\ news & 0.29  & 0.29  & 0.30  & 0.31  & 0.29  & 0.29  & 0.29 \\
$\sigma_\mu$  & invest.     & 18.63 & \textbf{18.73} & \textbf{19.13} & \textbf{19.63} & 18.63 & 18.63 & 18.63 \\
$\sigma_p$    & price m-up  & 0.16  & 0.16  & 0.16  & 0.16  & 0.16  & 0.16  & 0.16 \\
$\sigma_w$    & wage m-up   & 0.44  & 0.44  & 0.44  & 0.44  & 0.44  & 0.43  & 0.42 \\
$\sigma_{mp}$ & policy      & 0.38  & 0.38  & 0.38  & 0.38  & 0.38  & 0.38  & 0.38 \\
$\sigma_g$    & gov.       & 0.37  & 0.37  & 0.37  & 0.37  & 0.37  & 0.37  & 0.37 \\
\bottomrule
\end{tabular}
\vspace{0.3cm}
\begin{minipage}{\textwidth}
\justify
{\small \textit{Notes:} $\gamma_0^{med}$ denotes the posterior mean parameter vector under DE. KL denotes $KL_{ff}(\gamma_0^{med}, \gamma_c)$ with $\gamma_0^{med}$  corresponding to the benchmark specification. Panel~(a) allows all parameters to vary, while panel~(b) fixes $\sigma_\mu$ at its baseline value. The bold value signifies the binding constraint. Values are rounded to two decimal places except for~$\phi_x$.}
\end{minipage}
\end{table}

\begin{table}[h!]
\centering
\caption{KL and empirical distances between $\gamma_c$ and $\gamma_0$, HSY (2024) model}
\label{tab:kl_distances}
\vspace*{0.2cm}
\begin{tabular}{lccccccc}
\hline
& \multicolumn{3}{c}{(a) All parameters can vary} & \multicolumn{3}{c}{(b) $\sigma_\mu$ fixed} \\
\cmidrule(lr){2-4} \cmidrule(lr){5-7}
& $c=0.1$ & $c=0.5$ & $c=1.0$ & $c=0.1$ & $c=0.5$ & $c=1.0$ \\
\hline
KL & 7.96e-06 & 1.95e-04 & 7.63e-04 & 1.40e-05 & 2.90e-04 & 9.35e-04 \\
T=80 & 0.0543 & 0.0743 & 0.1058 & 0.0548 & 0.0748 & 0.1007 \\
T=150 & 0.0558 & 0.0839 & 0.1310 & 0.0568 & 0.0867 & 0.1283 \\
T=200 & 0.0566 & 0.0898 & 0.1473 & 0.0579 & 0.0942& 0.1464 \\
T=1000 & 0.0650 & 0.1591 & 0.3547 & 0.0694 & 0.1857 & 0.3844 \\
\hline
\end{tabular}
\par
\vspace*{0.3cm}
\begin{minipage}{1.02\textwidth}
\justify
{\small Note: KL denotes $KL_{ff}(\gamma_0^{med}, \gamma_c)$ with $\gamma_0^{med}$ corresponding to the posterior means under DE. The empirical distance measure equals $p_{ff}(\gamma_0^{med}, \gamma_c, 0.05, T)$, where T is specified in the last four rows of the table. Panel (a) shows results when all parameters can vary. Panel (b) shows results when $\sigma_\mu$ is fixed at its baseline value. $\kappa_p=(\epsilon_p-1)/\psi_p$, $\kappa_w=(\omega\epsilon_w)/\psi_w$. I follow \cite{l2024incorporating} and calibrate $\epsilon_p=\epsilon_w=6, \omega=1.$}
\end{minipage}
\end{table}

The model is driven by seven structural shocks: a total productivity shock, a noise shock to the signal about permanent productivity growth, a marginal efficiency of investment shock, price and wage markup shocks, and monetary and fiscal policy shocks. To estimate the model, \citet{l2024incorporating} use 12 observables: output growth, consumption growth, investment growth, wages, employment, inflation, the federal funds rate, and one-period-ahead forecasts of output growth, consumption growth, investment growth, inflation, and the interest rate. As noted by \citet{qu2023using}, however, global identification analysis requires the spectrum to be nonsingular. To satisfy this requirement, I work with log deviations from steady state for output, consumption, investment, wages, and employment, rather than their growth rates. Together with inflation and the interest rate, this yields seven observables in total. In the estimation stage, I also include five measurement errors, as in \citet{l2024incorporating}, to equate the number of shocks and observables. These measurement errors are excluded from the identification analysis, which is conducted using only the seven observables and the seven structural shocks. Appendix~\ref{sec: App_C.1} provides a detailed discussion of the modified information friction block together with the full set of model equations. As in the small-scale model, I take the posterior mean from Bayesian estimation as the benchmark parameter vector, denoted $\gamma_0^{med}$. Since the information friction block is modified, I re-estimate the medium-scale model using standard MCMC sampling.\footnote{I adopt the same priors as in \cite{l2024incorporating}. The resulting posterior distribution matches closely that reported in Table~1 of \cite{l2024incorporating}, with the only notable difference being the variance of the noise shock. This is an expected outcome since the signal now relates to the growth rate rather than the level of permanent productivity. The full posterior results are reported in Section~\ref{sec: App_C.2}.}

The global identification results are presented in Table \ref{tab:kl_params} and \ref{tab:kl_distances}\footnote{Local identification at $\gamma_0^{med}$ has been verified but is not reported here for brevity.}. When all parameters are allowed to vary, there is no evidence of observational equivalence. However, identification remains challenging with small sample sizes. The most problematic parameter is the shock variance $\sigma_\mu$, which exhibits the weakest identification among parameters\footnote{Under RE, it appears to be the second weakest identified parameter. Results are shown in the Online Appendix.}. 
To assess whether the identification results are driven by potentially misspecified low frequency components, I also conduct the exercise restricting attention to business cycle frequencies, which are the primary target of DSGE models. The results are broadly similar to the full frequency case and are reported in the Online Appendix.

\subsection{Observational Equivalence and the Role of Frictions} 
\label{sec: OE and Frictions Medium Scale}

Having established global identification in Section~\ref{Identification_med_FD}, I now examine observational equivalence and the role of frictions in the medium-scale model. In particular, I study whether DE and RE models are observationally equivalent in generating the dynamics of output, consumption, investment, real wages, labor, inflation, and the nominal interest rate; which frictions the RE model relies on most heavily when mimicking the DE benchmark; and which frictions are most important for generating the benchmark dynamic properties under DE.

The last two questions are motivated by \citet{l2024incorporating}, who show in a small-scale setting that DE and price rigidity interact to generate excess output volatility. Absent nominal rigidity, in a frictionless RBC model, the standard deviation of output can in fact be lower under DE than under RE. Whether other frictions in a medium-scale specification are important for generating the benchmark DE dynamics and interact similarly with DE is an open question that the identification strategy is well suited to address.

I address these issues by first allowing all RE parameters to vary over the feasible space, as in the small-scale analysis in Section~\ref{sec: DE_RE_small}. I then reduce individual frictions one at a time in the RE and DE models, re-optimizing the remaining parameters to minimize the KL divergence from the DE benchmark.

\subsubsection{Detecting observational equivalence between DE and RE } \label{sec: 4.2.1}
I begin by searching for the closest RE model in terms of KL criterion to the DE benchmark when all structural parameters are free to adjust. Taking the spectrum of the benchmark DE model at its posterior mean, $\gamma_0^{\mathrm{med}}$, I minimize the KL divergence over the feasible parameter bounds reported in the note to Table~\ref{tab: friction_comparison_DERE}.  The first two columns of Table~\ref{tab: friction_comparison_DERE} report the benchmark DE parameters and the corresponding closest RE parameters. Despite full freedom over all 25 structural parameters, the minimization yields no near observational equivalence: the empirical distance equals 0.99 at $T=80$ and 1.00 at $T=150$, indicating that the two models remain statistically distinguishable at empirically relevant sample sizes. The DE model therefore generates dynamics across the seven observation variables that no RE parameterization can replicate, even in the presence of a rich set of nominal, real, and information frictions.

To compensate for the absence of DE, the closest RE specification relies less on internal propagation and more on flexible adjustment, elastic labor supply and a dovish monetary policy rule. Relative to the DE benchmark, it features weaker habit formation, more flexible prices and wages, lower investment adjustment cost and more volatile markup shocks. The inverse Frisch elasticity declines substantially, from 3.71 to 2.41, and the signal noise parameter $\sigma_s$ also falls, from 0.29 to 0.18. Taken together, these shifts reveal a coherent compensating pattern: the closest RE specification moves toward more flexible adjustment and stronger contemporaneous responses to narrow the KL divergence from the DE benchmark.

\begin{figure}[h!]
\centering
\caption{Impulse responses to a government spending shock: DE baseline vs.\ RE (medium-scale model)}
\makebox[\textwidth][c]{%
    \includegraphics[width=1.2\textwidth, trim=30 10 30 10, clip]{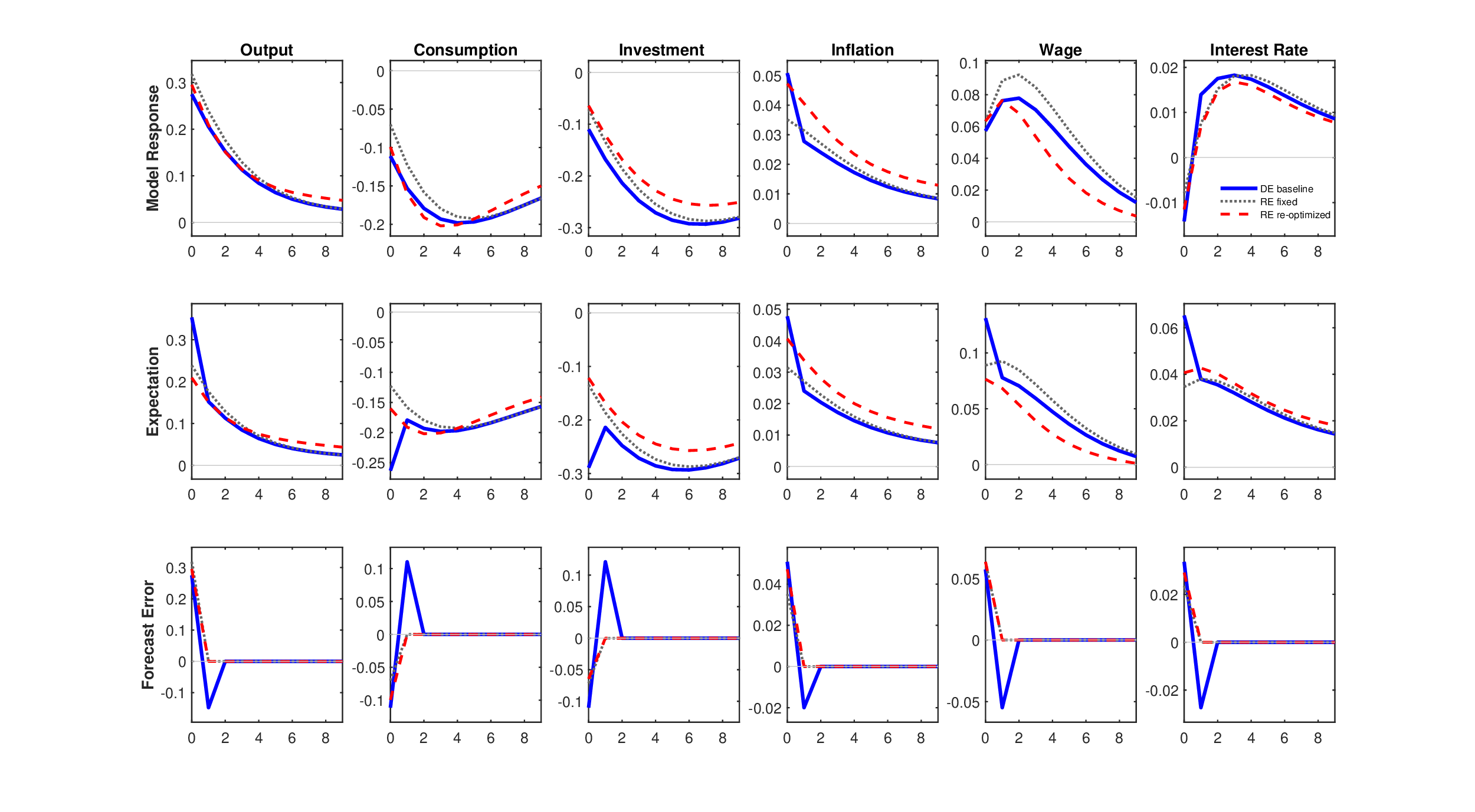}
}
\label{fig:irf_gov_DERE_unconstr}
\vspace{-10pt}
\begin{minipage}{1.02\textwidth}
\justify
{\small \textit{Note:} This figure displays impulse responses to a one-standard-deviation positive government spending shock. The top row reports the responses of output, consumption, investment, inflation, the real wages, and the subjective real interest rate. The middle row reports the corresponding one-period-ahead diagnostic expectations, $E_t^{\theta}[\cdot]$. The bottom row reports the corresponding forecast errors, $\eta_t(\cdot) \equiv x_t - E_{t-1}^{\theta}[\cdot]$. For the wages and interest-rate panels, the middle and bottom rows report expectations and forecast errors for the nominal wages and the nominal interest rate, respectively. The blue solid lines correspond to the DE baseline evaluated at the posterior mean $\gamma_0^{\mathrm{med}}$. The gray dotted lines correspond to the RE model evaluated at the same parameter values with $\theta=0$ (RE fixed). The red dashed lines correspond to the RE model with parameters chosen to minimize the KL divergence from the DE baseline (RE re-optimized). The horizontal axis is measured in quarters.}
\end{minipage}
\end{figure}

\begin{figure}[h!]
\centering
\caption{Impulse responses to a monetary policy shock: DE baseline vs.\ RE (medium-scale model)}
\makebox[\textwidth][c]{%
    \includegraphics[width=1.2\textwidth, trim=30 10 30 10, clip]{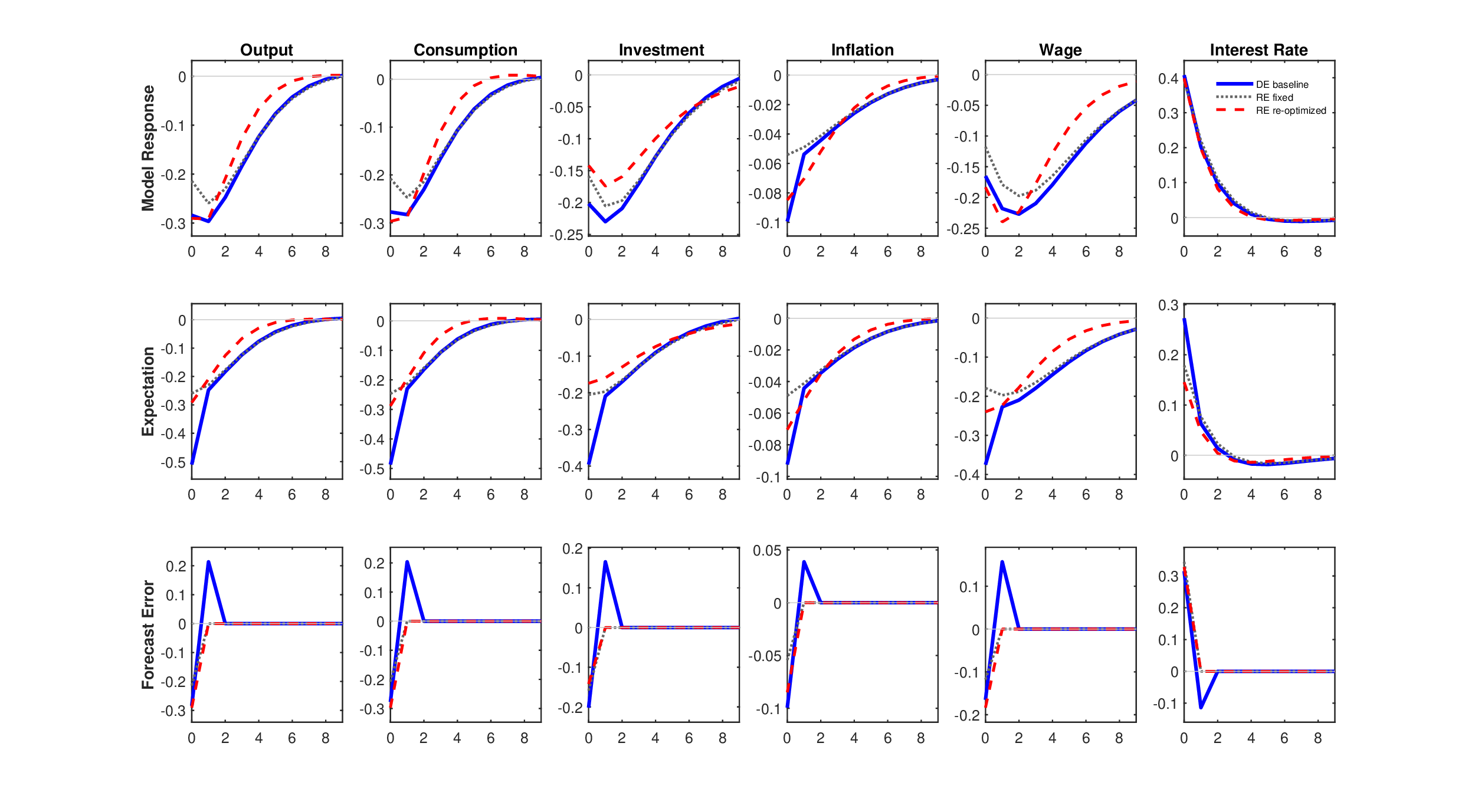}
}
\label{fig:irf_mp_DERE_unconstr}
\vspace{-10pt}
\begin{minipage}{1.02\textwidth}
\justify
{\small \textit{Note:} This figure displays impulse responses to a one-standard-deviation contractionary monetary policy shock. The top row reports the responses of output, consumption, investment, inflation, the real wages, and the subjective real interest rate. The middle row reports the corresponding one-period-ahead diagnostic expectations, $E_t^{\theta}[\cdot]$. The bottom row reports the corresponding forecast errors, $\eta_t(\cdot) \equiv x_t - E_{t-1}^{\theta}[\cdot]$. For the wages and interest-rate panels, the middle and bottom rows report expectations and forecast errors for the nominal wages and the nominal interest rate, respectively. The blue solid lines correspond to the DE baseline evaluated at the posterior mean $\gamma_0^{\mathrm{med}}$. The gray dotted lines correspond to the RE model evaluated at the same parameter values with $\theta=0$ (RE fixed). The red dashed lines correspond to the RE model with parameters chosen to minimize the KL divergence from the DE baseline (RE re-optimized). The horizontal axis is measured in quarters.}
\end{minipage}
\end{figure}

I now examine impulse responses to government spending and monetary policy shocks to provide intuition for which aspects of the DE dynamics these adjustments can and cannot reproduce. Figures~\ref{fig:irf_gov_DERE_unconstr} and~\ref{fig:irf_mp_DERE_unconstr} compare the DE benchmark with two RE specifications: the RE model evaluated at the DE posterior mean with $\theta=0$, and the closest RE model in terms of the KL criterion. In each figure, the top row shows the responses of output, consumption, investment, inflation, the real wages, and the subjective real interest rate, while the middle and bottom rows report the associated one-period-ahead expectations and forecast errors (with the wages and interest rate in nominal terms). The fixed-parameter RE specification isolates the direct effect of removing expectation extrapolation while holding all other parameters at their DE benchmark values, whereas the closest RE model shows how much of that gap can be closed through parameter re-optimization.

For a positive government spending shock, the closest RE model still fails to reproduce key features of the DE benchmark. In the small-scale model, as documented in Section~\ref{sec: DE_RE_small}, the largest discrepancy is in the subjective real rate: unlike under DE, the RE response is monotonic and displays no reversal. In the medium-scale model, the subjective real rate in the closest RE model also exhibits an initial decline followed by a reversal. This indicates that, with rich frictions, the closest RE model can partially mimic the real rate dynamics generated by DE. However, the closest RE response has a smaller impact decline and a weaker subsequent reversal than the DE benchmark.

Figure~\ref{fig:irf_gov_DERE_unconstr} shows three further patterns. First, for output and consumption, parameter re-optimization helps the closest RE model (red line) match the initial impact of the DE benchmark (blue line) more closely than the fixed-parameter RE model (gray line), but at the cost of weaker persistence. Real wages display a related but distinct discrepancy: the closest RE model matches the initial response relatively well but fails to reproduce the delayed peak and subsequent persistence under DE. Second, both RE specifications understate the crowding out of investment under DE throughout the horizon, and parameter re-optimization does not close this gap. Third, for inflation, the closest RE model comes closer to the DE impact response than the fixed-parameter RE model but overshoots the DE benchmark at longer horizons. The DE inflation response also displays a distinctive kink: a sharper impact response followed by faster initial decay. Intuitively, extrapolative expectations amplify inflation on impact and then dissipate, producing the kink, whereas the closest RE response, lacking this term, adjusts more gradually. The middle and bottom rows show that these differences originate in expectations. DE agents extrapolate the fiscal expansion on impact, and the resulting forecast errors display the characteristic pattern of overreaction followed by reversal.

For a contractionary monetary policy shock, shown in Figure~\ref{fig:irf_mp_DERE_unconstr}, the subjective real rate channel plays a more limited role in distinguishing DE from RE since the shock directly moves the nominal interest rate. Indeed, the subjective real rate response is nearly identical across all three specifications. The main discrepancy is instead an impact-persistence trade-off: the closest RE model can match either the impact magnitude or the persistence of the DE responses, but not both jointly. 
More specifically, for output, consumption, and real wages, the closest RE model matches the initial contraction under DE relatively well but exhibits weaker persistence. For investment, it understates the impact contraction but tracks the recovery under DE closely. For inflation, it comes closer on impact but overshoots the DE path at longer horizons, again smoothing out the distinctive DE kink seen under the government spending shock. By contrast, the fixed-parameter RE model, evaluated at the DE posterior mean with $\theta=0$, matches the persistence of the DE benchmark more closely because the two specifications share the same habit formation, nominal rigidities, and shock persistence. However, it fails to reproduce the larger initial impact under DE.

\subsubsection{Restricting frictions in RE model}\label{sec: 4.2.2}
\begin{table}[h!]
\centering
\caption{The closest RE models with constrained real and nominal frictions}
\begin{small}
\begin{tabular}{lcccccccc}
\hline
  & $\gamma_0^{\mathrm{med}}$ & $\theta = 0$ & $\sigma_s=0.1$ & $\kappa_p=1$ & $\kappa_w=1$ & $h=0.1$ & $\frac{\chi''(1)}{\chi'(1)}=1$ & $S''=1$ \\
\hline
KL & -- & 0.1137 & 0.1152 & 0.8218 & 3.8910 & 0.6135 & 1.3356 & 1.2214 \\
$T=80$ & -- & 0.9934 & 0.9938 & 1.0000 & 1.0000 & 1.0000 & 1.0000 & 1.0000 \\
$T=150$ & -- & 1.0000 & 1.0000 & 1.0000 & 1.0000 & 1.0000 & 1.0000 & 1.0000 \\[3pt]
$\theta$ & 0.72 & 0.00 & 0.00 & 0.00 & 0.00 & 0.00 & 0.00 & 0.00 \\
$\alpha$ & 0.13 & 0.14 & 0.14 & 0.14 & 0.13 & 0.13 & 0.15 & 0.16 \\
$h$ & 0.72 & 0.56 & 0.56 & 0.57 & 0.40 & -- & 0.34 & 0.54 \\
$\frac{\chi''(1)}{\chi'(1)}$ & 5.09 & 4.93 & 4.72 & 4.66 & 6.60 & 5.39 & -- & 5.61 \\
$\kappa_p$ & 0.04 & 0.07 & 0.07 & -- & 0.18 & 0.05 & 0.11 & 0.12 \\
$\kappa_w$ & 0.01 & 0.04 & 0.04 & 0.09 & -- & 0.05 & 0.19 & 0.05 \\
$\nu$ & 3.71 & 2.41 & 2.39 & 2.30 & 2.30 & 2.30 & 2.30 & 2.48 \\
$S''$ & 6.93 & 6.39 & 6.26 & 5.90 & 5.90 & 6.27 & 5.90 & -- \\
$\rho_R$ & 0.58 & 0.54 & 0.55 & 0.63 & 0.49 & 0.45 & 0.15 & 0.54 \\
$\phi_\pi$ & 1.54 & 1.32 & 1.33 & 2.08 & 1.96 & 1.22 & 1.54 & 1.53 \\
$\phi_x$ & 0.006 & 0.003 & 0.003 & 0.007 & 0.017 & 0.002 & 0.003 & 0.011 \\
$\rho$ & 0.85 & 0.81 & 0.82 & 0.69 & 0.81 & 0.88 & 0.54 & 0.76 \\
$\rho_\mu$ & 0.31 & 0.31 & 0.31 & 0.24 & 0.17 & 0.27 & 0.12 & 0.10 \\
$\rho_p$ & 0.88 & 0.90 & 0.90 & 0.92 & 0.92 & 0.89 & 0.89 & 0.90 \\
$\phi_p$ & 0.58 & 0.60 & 0.60 & 0.10 & 0.46 & 0.66 & 0.23 & 0.48 \\
$\rho_w$ & 0.99 & 0.99 & 0.99 & 0.99 & 0.99 & 0.99 & 0.99 & 0.99 \\
$\phi_w$ & 0.54 & 0.42 & 0.44 & 0.13 & 0.10 & 0.44 & 0.10 & 0.41 \\
$\rho_{mp}$ & 0.03 & 0.03 & 0.02 & 0.01 & 0.01 & 0.01 & 0.56 & 0.01 \\
$\rho_g$ & 0.94 & 0.94 & 0.95 & 0.94 & 0.93 & 0.92 & 0.98 & 0.96 \\
$\sigma_a$ & 1.43 & 1.46 & 1.45 & 1.46 & 1.46 & 1.43 & 1.81 & 1.55 \\
$\sigma_s$ & 0.29 & 0.18 & -- & 0.26 & 0.24 & 0.10 & 0.10 & 0.10 \\
$\sigma_\mu$ & 18.63 & 18.55 & 18.15 & 18.64 & 20.32 & 19.21 & 22.00 & 15.00 \\
$\sigma_p$ & 0.16 & 0.22 & 0.22 & 1.31 & 0.30 & 0.20 & 0.18 & 0.27 \\
$\sigma_w$ & 0.44 & 0.64 & 0.65 & 0.95 & 3.00 & 0.67 & 1.56 & 0.82 \\
$\sigma_{mp}$ & 0.38 & 0.38 & 0.38 & 0.40 & 0.45 & 0.39 & 0.55 & 0.39 \\
$\sigma_g$ & 0.37 & 0.37 & 0.37 & 0.37 & 0.37 & 0.37 & 0.45 & 0.37 \\
\hline
\end{tabular}
\end{small}
\par
\vspace*{0.3cm}
\begin{minipage}{\textwidth}
\justify
{\small \textit{Note:} The first column reports the posterior mean of the benchmark DE model, $\gamma_0^{\mathrm{med}}$. The second column ($\theta = 0$) reports the unrestricted RE model that minimizes the KL divergence with $\theta = 0$ and all other parameters free to vary within the parameter bounds ${\gamma} \in \{[0.1,\; 0.1,\; 3.6,\; 0.01,\; 0.001,\; 2.3,\; 
5.9,\; 0.1,\; 1.01,\; 0.001,\; 0.1,\; 0.1,\; 0.1,\; 0.1,\; 0.1,\; 0.1,\;0.01,\; 0.1,\; 1,\; 0.1,\\\; 15,\; 0.1,\; 0.1,\; 0.1,\; 0.1 \big], \big[2,\; 0.99,\; 6.6,\; 0.5,\; 0.5,\; 5.1,\; 
8.0,\; 0.99,\; 3,\; 0.99,\; 0.999,\; 0.999,\; 0.999,\; 0.999,\; \\
0.999,\; 0.999,\; 
0.9,\; 0.99,\; 3,\; 3,\; 22,\; 3,\; 3,\; 3,\; 3 ]\}$. Columns 3-8 report RE parameter vectors that minimize the KL divergence subject to the friction restriction indicated in each column header. KL equals $\mathrm{KL}_{ff}(\gamma_0^{\mathrm{med}}, \gamma^{\mathrm{RE}})$. $\kappa_p = (\epsilon_p - 1)/\psi_p$ and $\kappa_w = (\omega \epsilon_w)/\psi_w$ denote the slopes of the price and wage Phillips curves, with $\epsilon_p = \epsilon_w = 6$ and $\omega = 1$ following \citet{l2024incorporating}. All values rounded to two decimal places, except $\phi_x$.}
\end{minipage}
\label{tab: friction_comparison_DERE}
\end{table}

This subsection identifies which structural frictions are most important for matching the DE benchmark spectrum with a RE model. To do so, I reduce each friction in turn to a value substantially below its original level, while re-optimizing the remaining parameters to minimize the KL divergence from the DE benchmark. 
Columns 3-8 of Table~\ref{tab: friction_comparison_DERE} report the resulting KL divergences. 

Constraining wage rigidity produces the largest deviation, with a KL divergence of 3.89, followed by capital utilization costs at 1.34 and investment adjustment costs at 1.22. Price rigidity, at 0.82, and habit formation, at 0.61, occupy a middle tier, while constraining the signal noise variance has the smallest effect, at 0.12.
Appendix~\ref{sec: App_C.3} reports the corresponding exercise using the RE posterior mean as the benchmark. Under the RE benchmark, reducing investment adjustment costs results in the largest increase in KL divergence at 1.24, followed by capital utilization at 1.16, price rigidity at 0.97, habit formation at 0.60, wage rigidity at 0.54, and signal noise at 0.01. The KL divergence associated with constraining wage rigidity therefore rises sharply from 0.54 under the RE benchmark to 3.89 under the DE benchmark, whereas the corresponding increases for the other frictions are broadly similar across the two benchmarks. This comparison shows that the prominent role of wage rigidity is not a generic feature of the medium-scale model itself, but is specific to matching the DE benchmark spectrum.

\subsubsection{Restricting frictions in the DE model}\label{sec: 4.2.3}

This subsection extends the constrained-friction exercise of Section~\ref{sec: 4.2.2} by freeing the diagnosticity parameter $\theta$ to adjust. Each structural friction is constrained in turn to a value substantially below its DE posterior mean, and the KL divergence from the benchmark spectrum is minimized over the remaining parameters, with $\theta$ allowed to vary over $[0.1,2]$. This exercise identifies which frictions remain most important for matching the benchmark DE spectrum when diagnostic distortion itself is allowed to adjust. Furthermore, comparison with the corresponding fixed-$\theta$ results reveals heterogeneity in how DE interacts with structural frictions to generate the benchmark observable dynamics.

\begin{table}[h!]
\centering
\caption{The closest DE models with constrained real and nominal frictions}
\begin{tabular}{lccccccc}
\hline
 & $\gamma_0^{med}$ & $\sigma_s=0.1$ & $\kappa_p=1$ & $\kappa_w=1$ & $h=0.1$ & $\frac{\chi''(1)}{\chi'(1)}=1$ & $S''=1$ \\
\hline
KL & -- & 0.0066 & 0.8019 & 3.8574 & 0.6570 & 0.8910 & 0.4103 \\
$T=80$ & -- & 0.2730 & 1.0000 & 1.0000 & 1.0000 & 1.0000 & 1.0000 \\
$T=150$ & -- & 0.4092 & 1.0000 & 1.0000 & 1.0000 & 1.0000 & 1.0000 \\
$\theta$ & 0.72 & 0.69 & 0.24 & 0.22 & 0.10 & 2.00 & 1.29 \\
$\alpha$ & 0.13 & 0.14 & 0.14 & 0.13 & 0.13 & 0.15 & 0.15 \\
$h$ & 0.72 & 0.72 & 0.64 & 0.45 & -- & 0.75 & 0.77 \\
$\frac{\chi''(1)}{\chi'(1)}$ & 5.09 & 4.49 & 4.81 & 6.60 & 4.98 & -- & 3.97 \\
$\kappa_p$ & 0.04 & 0.04 & -- & 0.15 & 0.04 & 0.04 & 0.05 \\
$\kappa_w$ & 0.01 & 0.01 & 0.08 & -- & 0.05 & 0.08 & 0.01 \\
$\nu$ & 3.71 & 3.56 & 2.30 & 2.30 & 2.30 & 2.30 & 5.10 \\
$S''$ & 6.93 & 6.63 & 5.90 & 5.90 & 6.09 & 5.90 & -- \\
$\rho_R$ & 0.58 & 0.59 & 0.66 & 0.46 & 0.45 & 0.46 & 0.59 \\
$\phi_\pi$ & 1.54 & 1.59 & 2.29 & 2.00 & 1.25 & 1.68 & 1.92 \\
$\phi_x$ & 0.006 & 0.007 & 0.008 & 0.018 & 0.002 & 0.002 & 0.020 \\
$\rho$ & 0.85 & 0.86 & 0.75 & 0.88 & 0.88 & 0.70 & 0.82 \\
$\rho_\mu$ & 0.31 & 0.32 & 0.24 & 0.17 & 0.26 & 0.12 & 0.10 \\
$\rho_p$ & 0.88 & 0.88 & 0.93 & 0.91 & 0.89 & 0.92 & 0.85 \\
$\phi_p$ & 0.58 & 0.58 & 0.10 & 0.48 & 0.66 & 0.47 & 0.46 \\
$\rho_w$ & 0.99 & 0.99 & 0.99 & 0.99 & 0.99 & 0.99 & 0.99 \\
$\phi_w$ & 0.54 & 0.56 & 0.17 & 0.10 & 0.43 & 0.10 & 0.56 \\
$\rho_{mp}$ & 0.03 & 0.01 & 0.01 & 0.01 & 0.01 & 0.15 & 0.04 \\
$\rho_g$ & 0.94 & 0.94 & 0.94 & 0.93 & 0.92 & 0.98 & 0.95 \\
$\sigma_a$ & 1.43 & 1.38 & 1.42 & 1.45 & 1.41 & 1.51 & 1.33 \\
$\sigma_s$ & 0.29 & -- & 0.30 & 0.10 & 0.10 & 0.33 & 0.10 \\
$\sigma_\mu$ & 18.63 & 17.73 & 18.47 & 20.37 & 18.74 & 22.00 & 15.00 \\
$\sigma_p$ & 0.16 & 0.16 & 1.34 & 0.28 & 0.20 & 0.10 & 0.15 \\
$\sigma_w$ & 0.44 & 0.48 & 0.88 & 3.00 & 0.67 & 0.74 & 0.50 \\
$\sigma_{mp}$ & 0.38 & 0.38 & 0.41 & 0.48 & 0.39 & 0.42 & 0.40 \\
$\sigma_g$ & 0.37 & 0.37 & 0.37 & 0.37 & 0.37 & 0.45 & 0.37 \\
\hline
\end{tabular}
\label{tab: friction_comparison_DEDE}
\par
\vspace*{0.3cm}
\begin{minipage}{\textwidth}
\justify
{\small \textit{Note:} KL (row 2) and the empirical distance measures (rows 3--4) are defined as 
$KL_{ff}(\gamma_0^{med}, \gamma^{med})$ and $p_{ff}(\gamma_0^{med}, \gamma^{med}, 0.05, T)$, 
where the criterion is computed over the full frequency range. Each column reports the parameter 
vector $\gamma^{med}$ that minimizes the KL divergence from the benchmark DE parameter vector 
$\gamma_0^{med}$ (column 1), subject to the friction restriction indicated in the column header, 
while the diagnostic expectation parameter $\theta$ is left unrestricted. All parameter values 
are rounded to two decimal places, except $\phi_x$.}
\end{minipage}
\end{table}

\begin{figure}[h]
\centering
\caption{Impulse responses to a government spending shock: DE models with constrained wage rigidity}
\includegraphics[width=1\textwidth, trim=60 30 60 30, clip]{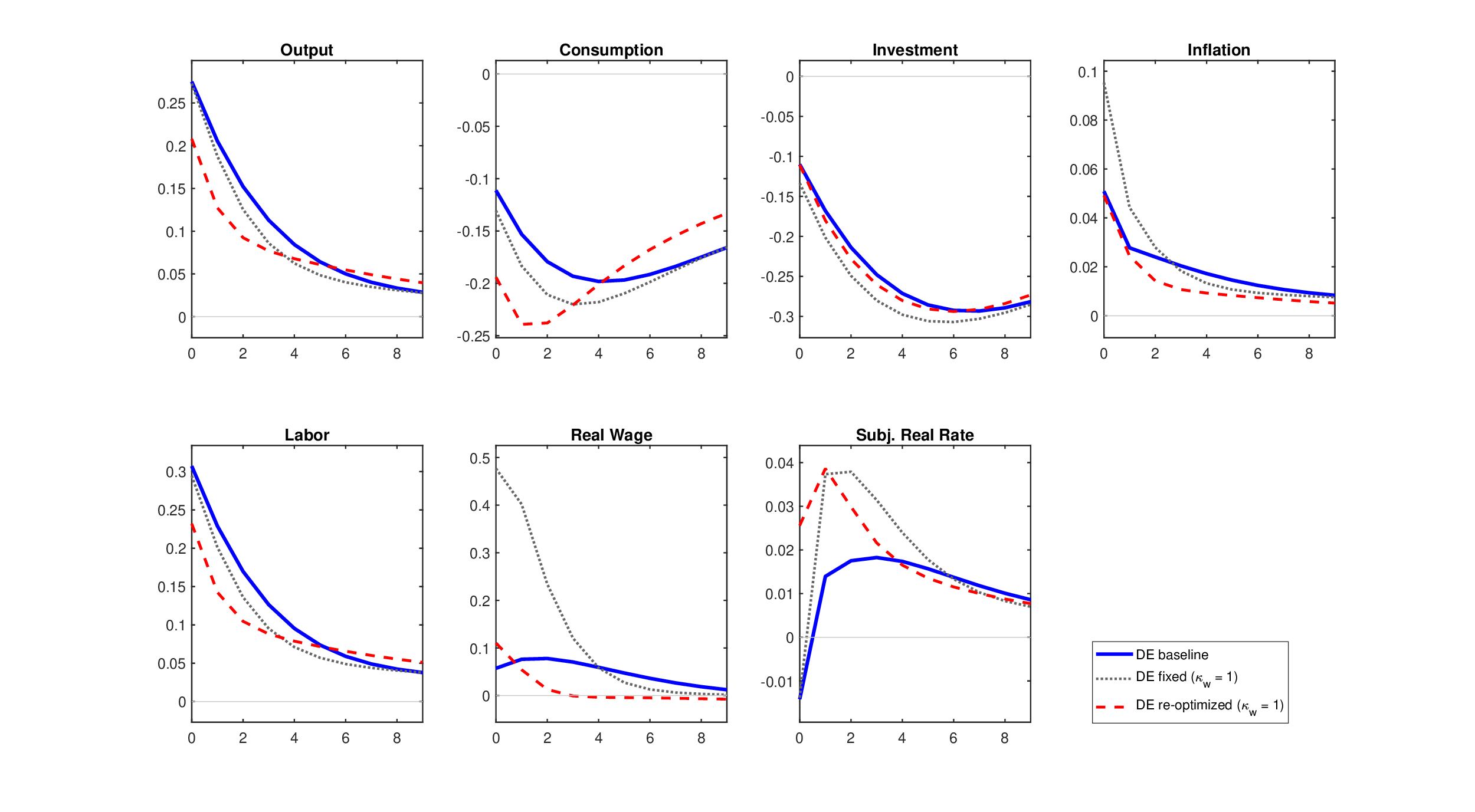}
\label{fig:irf_G_DEDE}

\vspace{-10pt}
\begin{minipage}{1\textwidth}
\justify
{\small \textit{Note:} This figure displays impulse responses to a one-standard-deviation positive government spending shock with wage rigidity constrained ($\kappa_w = 1$) in the DE model. The blue solid lines correspond to the DE baseline evaluated at the posterior mean $\gamma_0^{\mathrm{med}}$. The gray dotted lines correspond to the DE model with $\kappa_w = 1$ and all other parameters, including $\theta$, held at their DE baseline values (DE fixed). The red dashed lines correspond to the DE model with $\kappa_w = 1$ and all remaining parameters, including $\theta$, re-optimized to minimize the KL divergence from the DE baseline (DE re-optimized). The horizontal axis is in quarters.}
\end{minipage}
\end{figure}

\begin{figure}[h]
\centering
\caption{Impulse responses to a monetary policy shock: DE models with constrained wage rigidity}
\includegraphics[width=1\textwidth, trim=60 30 60 30, clip]{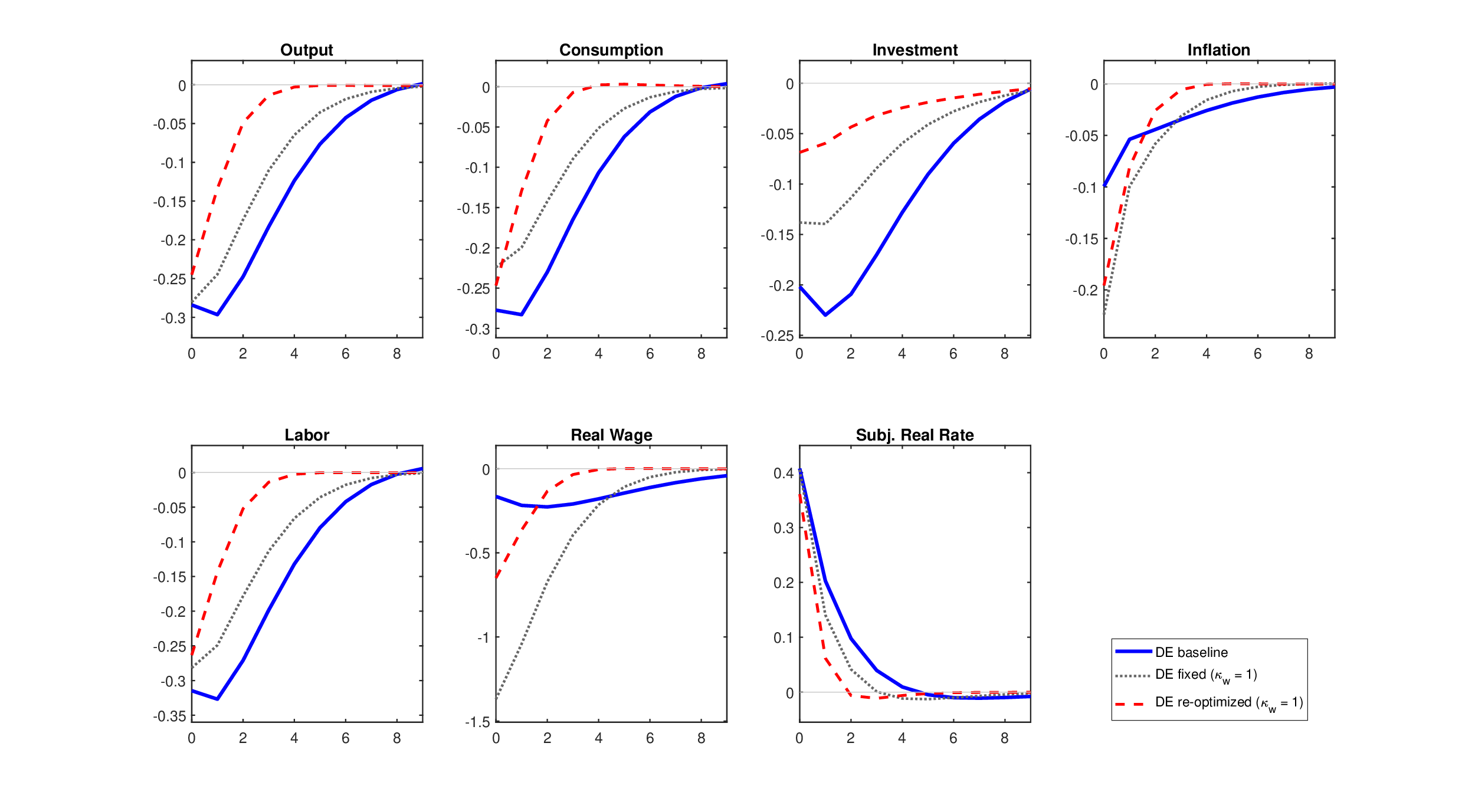}
\label{fig:irf_mp_DEDE}

\vspace{-10pt}
\begin{minipage}{1.02\textwidth}
\justify
{\small \textit{Note:} This figure displays impulse responses to a one-standard-deviation contractionary monetary policy shock with wage rigidity constrained ($\kappa_w = 1$) in the DE model. The blue solid lines correspond to the DE baseline evaluated at the posterior mean $\gamma_0^{\mathrm{med}}$. The gray dotted lines correspond to the DE model with $\kappa_w = 1$ and all other parameters, including $\theta$, held at their DE baseline values (DE fixed). The red dashed lines correspond to the DE model with $\kappa_w = 1$ and all remaining parameters, including $\theta$, re-optimized to minimize the KL divergence from the DE baseline (DE re-optimized). The horizontal axis is in quarters.}
\end{minipage}
\end{figure}

With $\theta$ additionally allowed to vary, Table~\ref{tab: friction_comparison_DEDE} shows that the friction ranking remains broadly consistent with that in Section~\ref{sec: 4.2.2}. Reducing wage rigidity still produces the largest KL divergence, at 3.86, followed by capital utilization cost, at 0.89, price rigidity, at 0.80, habit formation, at 0.66, and investment adjustment cost, at 0.41, while signal noise has only a small effect, at 0.007. Relative to the corresponding fixed-$\theta$ case, allowing $\theta$ to adjust reduces the KL divergence more substantially in the capital utilization cost and investment adjustment cost exercises than in the others. By contrast, in the wage rigidity, price rigidity, and habit formation exercises, the KL divergence remains largely unchanged.

For wage rigidity in particular, the optimizer compensation pattern closely mirrors that in the fixed-$\theta$ case. When wage rigidity is constrained, the KL minimizer raises the standard deviation of the wage markup shock to its upper bound of 3.00, increases the standard deviation of the MEI shock, relaxes price rigidity, and reduces habit formation. $\theta$ falls to 0.22, yet the KL divergence remains at 3.86, essentially unchanged from the corresponding KL fixed-$\theta$ value of 3.89. This shows that wage rigidity is essential for generating the distinctive benchmark dynamics under DE, and that these dynamics cannot be restored either by reallocating across other structural frictions or by adjusting $\theta$.

To interpret this result intuitively, it is useful to consider the IRFs and the structural equations together. Following a positive government spending shock, aggregate demand rises and, under DE, this increase is further amplified by extrapolative expectations. When wage rigidity is reduced while all other parameters are held at their DE benchmark values (gray line), real wages adjust more rapidly than in the DE benchmark (blue line). This faster wage adjustment feeds more quickly into firms' labor costs and hence into marginal cost, which enters the price Phillips curve. Inflation therefore rises more sharply on impact, so the gap between demand and supply closes more quickly than in the benchmark, leaving less scope for DE to generate additional amplification over time. As emphasized by \citet{l2024incorporating}, the extent to which aggregate demand can diverge from supply depends on the degree of nominal rigidities. Although the KL minimizer (red line) mitigates the sharp real wages response through re-optimization, it cannot restore the more persistent benchmark dynamics across most observables, with investment as the main exception. A similar pattern operates under a contractionary monetary policy shock, where lower wage rigidity speeds nominal adjustment and compresses the persistence of inflation and real activity relative to the DE benchmark.

As for capital utilization cost, reducing it induces parameter shifts in the closest DE model along multiple dimensions. When $\theta$ is additionally allowed to vary, it rises to its upper bound of 2.00, and the accompanying shifts in the remaining structural parameters are smaller than in the corresponding fixed-$\theta$ case. Habit formation, for example, remains close to its benchmark value at 0.75 rather than falling to 0.34, and the relaxation of price and wage rigidity is more modest. Shock-related parameters also adjust less aggressively: the standard deviation of the wage markup shock rises only to 0.74 rather than 1.56, and that of the TFP shock to 1.51 rather than 1.81. The resulting KL divergence falls to 0.89, substantially below the corresponding fixed-$\theta$ value of 1.34. This suggests that capital utilization cost contributes substantially to matching the benchmark dynamics on its own, but that DE can partially substitute for its role.

The remaining frictions are summarized briefly here. Under constrained price rigidity, the closest DE model exhibits a parameter adjustment pattern similar to that under constrained wage rigidity: the optimizer raises the standard deviation of the price markup shock from benchmark value 0.16 to 1.34, increases the monetary policy reaction coefficient from 1.54 to 2.29, and lowers $\theta$ to 0.24. Yet allowing $\theta$ to vary does little to close the KL divergence, with the KL falling only marginally from 0.82 to 0.80. Under constrained investment adjustment cost, by contrast, the closest DE model relies less on markup shock volatility than the closest RE model, while $\theta$ rises to 1.29; the KL divergence falls to 0.41, well below the corresponding fixed-$\theta$ value of 1.22, indicating partial substitutability between investment adjustment cost and diagnosticity. Under constrained habit formation, the closest DE model exhibits a parameterization similar to that of the closest RE model, with a more elastic labor supply and a more dovish monetary policy rule; when $\theta$ is allowed to vary, however, the optimizer pushes it to its lower bound, indicating essentially no compensation for the loss of habit persistence. Reducing signal noise while allowing $\theta$ to vary, the optimizer remains close to the benchmark: $\theta$ moves only from 0.72 to 0.69, and the main adjustment is a  reduction in the standard deviation of the MEI shock. The KL divergence falls from 0.12 to 0.007. This indicates that signal noise contributes little to the benchmark observable dynamics in the DE model.

\section{Conclusion}\label{sec: conclusion}
This paper shows that DE are identifiable in log-linearized DSGE models and generate observable dynamics that cannot be replicated by RE models. Using the frequency-domain framework of \citet{qu2017global}, I first study a small-scale model and show that the diagnosticity parameter $\theta$ is locally and globally identified at the posterior mean. Moreover, no RE parameterization can replicate the spectrum of output, inflation, and the nominal interest rate implied by the DE benchmark, although introducing $\theta$ weakens the identification of the shock variances.

The medium-scale analysis confirms these findings. DE remains globally identified, and no RE model replicates the spectrum of output, consumption, investment, real wages, labor, inflation, and the nominal interest rate, even when all remaining structural parameters are allowed to adjust. Impulse responses show that the closest RE model cannot jointly match the initial impact and the subsequent persistence of output, consumption, investment, and real wages responses to a monetary policy shock, nor the magnitude of the impact decline and reversal in the subjective real interest rate following a government spending shock. The distinctive kink in inflation dynamics, generated by extrapolative expectations, is also absent under RE. Among the structural frictions, wage rigidity emerges as by far the most important for matching the DE benchmark spectrum. 

Two avenues remain for future work. A natural extension is to study the identification of DE under equilibrium indeterminacy. \cite{lubik2003computing} show how to estimate DSGE models in the indeterminacy region by introducing sunspot shocks alongside structural shocks, while \citet{qu2017global} document cases in which parameters that are unidentified under determinacy become identifiable once indeterminacy is admitted. \citet{hirose2026behavioral} provide a useful starting point by incorporating DE into a model under indeterminacy and documenting a greater importance for sunspot shocks under DE. A second avenue is to examine whether including forecast variables as observables helps identify the shock variances, given that introducing $\theta$ weakens their identification in the present analysis. If so, this would further underscore the importance of survey data on expectations for the empirical evaluation of DE models.
	\clearpage
	\newpage
	\bibliographystyle{ecta}
	\bibliography{main}

@article{kahneman1972subjective,
	title={Subjective probability: A judgment of representativeness},
	author={Kahneman, Daniel and Tversky, Amos},
	journal={Cognitive psychology},
	volume={3},
	number={3},
	pages={430--454},
	year={1972},
	publisher={Elsevier}
}

@ARTICLE{Sims03,
	author = { Sims, C.},
	title = {Implications of Rational Inattention},
	journal = {Journal of Monetary Economics},
	year = {2003},
	volume = {50},
	pages = {665-690}
}

@article{bordalo2018diagnostic,
	title={Diagnostic expectations and credit cycles},
	author={Bordalo, Pedro and Gennaioli, Nicola and Shleifer, Andrei},
	journal={The Journal of Finance},
	volume={73},
	number={1},
	pages={199--227},
	year={2018},
	publisher={Wiley Online Library}
}

@article{l2024incorporating,
  title={Incorporating diagnostic expectations into the New Keynesian framework},
  author={L’Huillier, Jean-Paul and Singh, Sanjay R and Yoo, Donghoon},
  journal={Review of Economic Studies},
  volume={91},
  number={5},
  pages={3013--3046},
  year={2024},
  publisher={Oxford University Press UK}
}

@article{canova2009back,
  title={Back to square one: Identification issues in DSGE models},
  author={Canova, Fabio and Sala, Luca},
  journal={Journal of Monetary Economics},
  volume={56},
  number={4},
  pages={431--449},
  year={2009},
  publisher={Elsevier}
}

@article{qu2012identification,
  title={Identification and frequency domain quasi-maximum likelihood estimation of linearized dynamic stochastic general equilibrium models},
  author={Qu, Zhongjun and Tkachenko, Denis},
  journal={Quantitative Economics},
  volume={3},
  number={1},
  pages={95--132},
  year={2012},
  publisher={Wiley Online Library}
}

@article{iskrev2010local,
  title={Local identification in DSGE models},
  author={Iskrev, Nikolay},
  journal={Journal of Monetary Economics},
  volume={57},
  number={2},
  pages={189--202},
  year={2010},
  publisher={Elsevier}
}

@article{anderson2008solving,
  title={Solving linear rational expectations models: A horse race},
  author={Anderson, Gary S},
  journal={Computational Economics},
  volume={31},
  pages={95--113},
  year={2008},
  publisher={Springer}
}

@article{qu2017global,
  title={Global identification in DSGE models allowing for indeterminacy},
  author={Qu, Zhongjun and Tkachenko, Denis},
  journal={The Review of Economic Studies},
  volume={84},
  number={3},
  pages={1306--1345},
  year={2017},
  publisher={Oxford University Press}
}

@techreport{bordalo2021real,
	title={Real credit cycles},
	author={Bordalo, Pedro and Gennaioli, Nicola and Shleifer, Andrei and Terry, Stephen J},
	year={2021},
	institution={National Bureau of Economic Research}
}

@article{bordalo2021diagnostic,
	title={Diagnostic bubbles},
	author={Bordalo, Pedro and Gennaioli, Nicola and Kwon, Spencer Yongwook and Shleifer, Andrei},
	journal={Journal of Financial Economics},
	volume={141},
	number={3},
	pages={1060--1077},
	year={2021},
	publisher={Elsevier}
}

@article{bordalo2020overreaction,
	title={Overreaction in macroeconomic expectations},
	author={Bordalo, Pedro and Gennaioli, Nicola and Ma, Yueran and Shleifer, Andrei},
	journal={American Economic Review},
	volume={110},
	number={9},
	pages={2748--82},
	year={2020}
}

@article{bianchi2024diagnostic,
  title={Diagnostic business cycles},
  author={Bianchi, Francesco and Ilut, Cosmin and Saijo, Hikaru},
  journal={Review of Economic Studies},
  volume={91},
  number={1},
  pages={129--162},
  year={2024},
  publisher={Oxford University Press US}
}

@article{smets2007shocks,
  title={Shocks and frictions in US business cycles: A Bayesian DSGE approach},
  author={Smets, Frank and Wouters, Rafael},
  journal={American economic review},
  volume={97},
  number={3},
  pages={586--606},
  year={2007},
  publisher={American Economic Association}
}

@article{lubik2003computing,
  title={Computing sunspot equilibria in linear rational expectations models},
  author={Lubik, Thomas A and Schorfheide, Frank},
  journal={Journal of Economic dynamics and control},
  volume={28},
  number={2},
  pages={273--285},
  year={2003},
  publisher={Elsevier}
}

@article{cai2021online,
  title={Online estimation of DSGE models},
  author={Cai, Michael and Del Negro, Marco and Herbst, Edward and Matlin, Ethan and Sarfati, Reca and Schorfheide, Frank},
  journal={The Econometrics Journal},
  volume={24},
  number={1},
  pages={C33--C58},
  year={2021},
  publisher={Oxford University Press}
}

@article{herbst2014sequential,
  title={Sequential Monte Carlo sampling for DSGE models},
  author={Herbst, Edward and Schorfheide, Frank},
  journal={Journal of Applied Econometrics},
  volume={29},
  number={7},
  pages={1073--1098},
  year={2014},
  publisher={Wiley Online Library}
}

@article{schorfheide2008dsge,
  title={DSGE model-based estimation of the New Keynesian Phillips curve},
  author={Schorfheide, Frank},
  journal={FRB Richmond Economic Quarterly},
  volume={94},
  number={4},
  pages={397--433},
  year={2008}
}

@article{del2008forming,
  title={Forming priors for DSGE models (and how it affects the assessment of nominal rigidities)},
  author={Del Negro, Marco and Schorfheide, Frank},
  journal={Journal of Monetary Economics},
  volume={55},
  number={7},
  pages={1191--1208},
  year={2008},
  publisher={Elsevier}
}

@techreport{jones2021priors,
  title={Priors and the Slope of the Phillips Curve},
  author={Jones, Callum and Kulish, Mariano and Nicolini, Juan Pablo},
  year={2021},
  institution={JSTOR}
}

@article{hazell2022slope,
  title={The slope of the Phillips Curve: evidence from US states},
  author={Hazell, Jonathon and Herreno, Juan and Nakamura, Emi and Steinsson, J{\'o}n},
  journal={The Quarterly Journal of Economics},
  volume={137},
  number={3},
  pages={1299--1344},
  year={2022},
  publisher={Oxford University Press}
}

@book{gali2015monetary,
  title={Monetary policy, inflation, and the business cycle: an introduction to the new Keynesian framework and its applications},
  author={Gal{\'\i}, Jordi},
  year={2015},
  publisher={Princeton University Press}
}

@article{nakamura2014fiscal,
  title={Fiscal stimulus in a monetary union: Evidence from US regions},
  author={Nakamura, Emi and Steinsson, J{\'o}n},
  journal={American Economic Review},
  volume={104},
  number={3},
  pages={753--792},
  year={2014},
  publisher={American Economic Association 2014 Broadway, Suite 305, Nashville, TN 37203}
}

@inproceedings{planas2015slice,
  title={Slice sampling in Bayesian estimation of DSGE models},
  author={Planas, Christophe and Ratto, Marco and Rossi, Alessandro},
  booktitle={Conference paper presented at 11th DYNARE conference},
  year={2015}
}

@article{an2007bayesian,
  title={Bayesian analysis of DSGE models},
  author={An, Sungbae and Schorfheide, Frank},
  journal={Econometric reviews},
  volume={26},
  number={2-4},
  pages={113--172},
  year={2007},
  publisher={Taylor \& Francis}
}

@article{blanchard2013news,
  title={News, noise, and fluctuations: An empirical exploration},
  author={Blanchard, Olivier J and L'Huillier, Jean-Paul and Lorenzoni, Guido},
  journal={American Economic Review},
  volume={103},
  number={7},
  pages={3045--3070},
  year={2013},
  publisher={American Economic Association}
}

@article{komunjer2011dynamic,
  title={Dynamic identification of dynamic stochastic general equilibrium models},
  author={Komunjer, Ivana and Ng, Serena},
  journal={Econometrica},
  volume={79},
  number={6},
  pages={1995--2032},
  year={2011},
  publisher={Wiley Online Library}
}

@article{koop2013identification,
  title={On identification of Bayesian DSGE models},
  author={Koop, Gary and Pesaran, M Hashem and Smith, Ron P},
  journal={Journal of Business \& Economic Statistics},
  volume={31},
  number={3},
  pages={300--314},
  year={2013},
  publisher={Taylor \& Francis}
}

@article{kocikecki2023solution,
  title={A solution to the global identification problem in DSGE models},
  author={Koci{\k{e}}cki, Andrzej and Kolasa, Marcin},
  journal={Journal of Econometrics},
  volume={236},
  number={2},
  pages={105477},
  year={2023},
  publisher={Elsevier}
}

@article{qu2023using,
  title={Using arbitrary precision arithmetic to sharpen identification analysis for DSGE models},
  author={Qu, Zhongjun and Tkachenko, Denis},
  journal={Journal of Applied Econometrics},
  volume={38},
  number={4},
  pages={644--667},
  year={2023},
  publisher={Wiley Online Library}
}

@article{na2025overreaction,
  title={Overreaction and macroeconomic fluctuation of the external balance},
  author={Na, Seunghoon and Yoo, Donghoon},
  journal={Journal of Monetary Economics},
  volume={151},
  pages={103750},
  year={2025},
  publisher={Elsevier}
}

@techreport{bianchi2024smooth,
  title={Smooth diagnostic expectations},
  author={Bianchi, Francesco and Ilut, Cosmin L and Saijo, Hikaru},
  year={2024},
  institution={National Bureau of Economic Research}
}

@article{yin2023diagnostic,
  title={Diagnostic Expectations and Consumption Dynamics},
  author={Guo, Jinting and Luo, Yulei and Yin, Penghui},
  journal={Available at SSRN 5173669},
  year={2023}
}

@article{hirose2026behavioral,
  title={Behavioral Expectations Under Indeterminacy: An Empirical Evaluation CAMA Working Paper 2/2026 January 2026},
  author={Hirose, Yasuo and Yoo, Donghoon},
  year={2026}
}
	\clearpage
    \newpage

\appendix

\counterwithin{table}{section}
\counterwithin{figure}{section}

\renewcommand{\thetable}{\Alph{section}.\arabic{table}}
\renewcommand{\thefigure}{\Alph{section}.\arabic{figure}}

\section*{Appendix}
\addcontentsline{toc}{section}{Appendix}
\section{Diagnostic Expectations}
\label{app:DE_theory} \label{app: DE_theory}
This appendix provides the formal derivation of the DE framework summarized in Section~\ref{sec: DE_summary}. I present the distorted probability distribution underlying DE, derive its RE representation, and detail the solution method of \citet{l2024incorporating} for embedding DE into the stochastic difference equation (SDE) system of a log-linearized DSGE model. 

DE originates from the representativeness heuristic of \citet{kahneman1972subjective}. An attribute is perceived as representative of a group when its relative frequency is higher in that group than in alternative groups. Because individuals often assess likelihood via representativeness, they tend to overestimate the probability of attributes they perceive as representative. 

\cite{bordalo2018diagnostic} first formalized the DE for an exogenous economic state variable following AR(1) process. Assume the economic state at t is $\omega_t$ following an AR(1) process $%
\omega_t=\rho \omega_{t-1}+\varepsilon_t$, where $\varepsilon_t\sim N(0,\sigma_{\varepsilon}^2)$ and $%
\rho\in(0,1]$ is the persistent parameter. The more representative future state is the one more likely to occur under the realized
state $G\equiv \left\{\omega_t=\hat{\omega_t}\right\}$ than based on the referenced past $-G\equiv \left\{\omega_t=\rho\hat{\omega}_{t-1}\right\}$. Hence, the representativeness can be written as a division of the two conditional probability distributions
$\frac{f(\hat \omega_{t+1}|G_t)%
}{f(\hat \omega_{t+1}|-G_t)}$. When DE agents make their expectations, they have the true conditional expectation in mind but inflate the probability of the representative future state and deflate the less representative one. Therefore, the
diagnostic distribution (or the distorted pdf) of $\omega_{t+1}$ is defined as true distribution times the representative-distortion term
\begin{equation}
f_t^{\theta}(\hat \omega_{t+1})=f(\hat \omega_{t+1}|G_t)\left[\frac{f(\hat \omega_{t+1}|G_t)%
}{f(\hat \omega_{t+1}|-G_t)}\right]^{\theta}\cdot C,
\end{equation}
where $C$ is a constant ensuring $f_t^{\theta}$ integrate to 1 and $\theta$ measures the distortion severity. If $\theta=0$, then representative distortion shuts down, we are going back to the RE case. If $\theta>0$, the larger the $\theta$, the larger overweighting of the representative state. Denote the diagnostic expectation operator at time $t$ by $E_t^{\theta}$, it can be formally defined as 
\begin{equation}
E_t^{\theta}[\omega_{t+1}]=\int_{-\infty}^{\infty}\omega f_t^{\theta}(\omega)d\omega.
\end{equation}
Since $\omega_t$ follows an AR(1) process with $N(0, \sigma_\varepsilon^2)$ shocks, it 
is very crucial to point out that the diagnostic distribution is also normal. Thus DE has a RE representation\footnote{For proof, see the Internet Appendix of \cite{bordalo2018diagnostic}. It is also shown in the appendix that the property can easily expand to the case where $\omega_t$ follows a AR(N) process. }
\begin{equation}
E_t^{\theta}(\omega_{t+1})=E_t\omega_{t+1}+\theta[E_t\omega_{t+1}-E_{t-1}\omega_{t+1}].
\end{equation}
The RE representation also holds for the multivariate case \citep{l2024incorporating}. 

Although the original analysis of DE lies on autoregressive exogenous variables \citep{bordalo2018diagnostic}, studying the DE for endogenous variables is crucial for solving economic models like the DSGE model with DE agents. \cite{l2024incorporating} propose a solution method that solves a stochastic difference equation system combining both exogenous and endogenous variables. Suppose the SDE is
\begin{equation}
E_t^\theta[\boldsymbol{Fy}_{t+1}+\boldsymbol{G_1y}_t+\boldsymbol{Mx}_{t+1}+\boldsymbol{N_1x}_t]+\boldsymbol{G_2y}_t+\boldsymbol{Hy}_{t-1}+\boldsymbol{N_2x}_t=0
\end{equation}
where exogenous variables are stacked in a $(n\times 1)$ vector $\boldsymbol x_t$
following an AR(1) stochastic process, i.e., $\boldsymbol x_t=\boldsymbol{Ax}_{t-1}+\boldsymbol\nu_t$ and $\boldsymbol A$ is a diagonal matrix of persistence parameters, $\boldsymbol\nu_t\sim N(0,\Sigma_\nu)$; $\boldsymbol y_t $ is a $(m\times 1)$ vector of endogenous variables; $\boldsymbol F_{m\times m} , (\boldsymbol{G_1})_{m\times m} , (\boldsymbol{G_2})_{m\times m}, \boldsymbol H_{m\times m}, $ $(\boldsymbol{N_1})_{m\times n}  $\,\,and\,\,$  (\boldsymbol{N_2})_{m\times n}$ are matrices of parameters.

To write the RE representation for SDE combined with exogenous and endogenous variables, \cite{l2024incorporating} guess a solution
according to the extrapolative nature of DE, i.e., $\boldsymbol y_t=\boldsymbol{Py}_{t-1}+\boldsymbol{Qx}_t+\boldsymbol{R\nu}_t$. After verification, they show it indeed constitutes a solution for SDE.\footnote{For details, see the appendix of \cite{l2024incorporating}.} Note that the solution has a very good property in that it follows a multivariate normal distribution. Hence using the same technology as exogenous normal distributed variables, the DE-SDE has the following RE representation
\begin{align}\label{RErep}
    \boldsymbol{F}E_t[\boldsymbol y_{t+1}]+\boldsymbol G\boldsymbol y_t+\boldsymbol{Hy}_{t-1}+\boldsymbol{M}E_t[\boldsymbol x_{t+1}]+\boldsymbol N\boldsymbol x_t
          +\boldsymbol F\theta(E_t[\boldsymbol y_{t+1}]-E_{t-1}[\boldsymbol y_{t+1}])\\ \nonumber
          +\boldsymbol M\theta(E_t[\boldsymbol x_{t+1}]-E_{t-1}[\boldsymbol x_{t+1}])+\boldsymbol G_1\theta(\boldsymbol y_t-E_{t-1}[\boldsymbol y_t])+\boldsymbol N_1\theta(\boldsymbol x_t-E_{t-1}[\boldsymbol x_t])=0 
\end{align}
where $\boldsymbol G=\boldsymbol G_1+\boldsymbol G_2$, $\boldsymbol N=\boldsymbol N_1+\boldsymbol N_2$.

\section{A small-scale DSGE model} \label{sec: AppB}
\subsection{Analytical solution}  \label{sec: App_B.1}
An intuitive way of understanding why adding diagnostic distortion weakens the identification strength of the shock variance relatively is to examine the analytical solution. Below, I present the analytical solution using the guess-and-verify method. To simplify the analysis, I consider a two-equation system, which is the benchmark system without the monetary policy rule and with the interest rate being constant:
    \begin{align*}
         (1+\theta)E_t[\hat{y}_{t+1}]+(1+\theta)E_t[\hat{\pi}_{t+1}]-\hat{y}_t+\hat{g}_t+\theta\hat{\pi}_t-(1+\theta)E_t[\hat{g}_{t+1}],\nonumber\\
         =\theta E_{t-1} [\hat{y}_{t+1}]+\theta E_{t-1}[\hat{\pi}_{t+1}]+\theta E_{t-1}[\hat{\pi}_t] -\theta E_{t-1}[\hat{g}_{t+1}],\\ \nonumber
         \hat{\pi}_t=\beta(1+\theta)E_t [\hat{\pi}_{t+1}]-\beta \theta E_{t-1} [\hat{\pi}_{t+1}]+\kappa(\hat{y}_t-\hat{a}_t)-\kappa \psi \hat{g}_t,\\ \nonumber
        \hat{a}_t=\rho_a\hat{a}_{t-1}+\varepsilon_{a,t}, \quad \hat{g}_t=\rho_g\hat{g}_{t-1}+\varepsilon_{g,t}      
        \end{align*}
Guess the solution is in the form of
\begin{align*}
    \hat{y}_t=\alpha_{11}\hat{a}_{t-1}+\alpha_{12}\hat{g}_{t-1}+\mu_{11}\varepsilon_{a,t}+\mu_{12}\varepsilon_{g,t}, \quad
    \hat{\pi}_t=\alpha_{21}\hat{a}_{t-1}+\alpha_{22}\hat{g}_{t-1}+\mu_{21}\varepsilon_{a,t}+\mu_{22}\varepsilon_{g,t}
\end{align*}
Plugging in the guess solution, collecting terms and comparing coefficient yields

\begin{align*}
\alpha_{11} &= -\frac{\kappa\rho_a^2}{1 - \rho_a(1 + \beta + \kappa) + \beta \rho_a^2}, \qquad
\alpha_{12} = \frac{\rho_g \left[1 - \rho_g(1 + \beta + \kappa \psi) + \beta\rho_g^2\right]}{1 - \rho_g(1 + \beta + \kappa) + \beta \rho_g^2} \\
\alpha_{21} &= -\frac{\kappa (1 - \rho_a) \rho_a}{1 - \rho_a(1 + \beta + \kappa) + \beta \rho_a^2}, \qquad
\alpha_{22} = \frac{\kappa \rho_g (1 - \psi)(1 - \rho_g)}{1 - \rho_g(1 + \beta + \kappa) + \beta \rho_g^2}
\end{align*} 
\begin{align*}
\mu_{11} &=-\frac{\kappa \Big[ \rho_a 
   + \theta(1 - \kappa \rho_a) 
   + \beta \rho_a \theta^2 (1 - \rho_a)\Big]}
{[1 - \rho_a(1 + \beta + \kappa ) + \beta \rho_a^2](1-\kappa \theta)}, \qquad
       \mu_{21} &= -
\frac{\kappa\Big[(1 - \rho_a) + \rho_a\theta\,[\kappa+\beta (1-\rho_a)]\Big]}
{[1  - \rho_a(1 + \beta + \kappa ) + \beta \rho_a^2]\,(1-\kappa \theta )},\qquad
 \mu_{22} = \frac{(1+\theta)[\beta\alpha_{22} + \kappa(\alpha_{12}+\alpha_{22}-\rho_g)] 
+ \kappa(1 -\psi)}{1-\kappa\theta}.
\end{align*}
where $\alpha_{11}, \alpha_{12}, \alpha_{21}, \alpha_{22}$ are the same as in the RE case, while DE affects the shock terms $\mu_{11}, \mu_{12}, \mu_{21}, \mu_{22}.$. This result is in line with the theoretical predictions. As demonstrated by \cite{bordalo2018diagnostic}, DE induce overreaction in the dynamics of exogenous processes and, consequently, amplifies their effects on endogenous variables. Since the diagnosticity parameter $\theta$ operates exclusively through the transmission of shocks to endogenous variables, variations in shock variances can be partially offset by adjustments in $\theta$. The model-implied spectrum is therefore less sensitive to changes in shock variances, weakening their identification strength.

\subsection{Identification result for rational expectation model} \label{sec: App_B.2}
I use the replication code from \citet{qu2017global} to minimize $KL(\gamma_0^{RE}, \gamma^{RE})$ under the constraint $\|\gamma^{RE} - \gamma_0^{RE}\|_{\infty} \geq c$ for the benchmark RE model. The results are presented in Tables \ref{tab: optimizer_RE} and \ref{tab: KL_RE}. When all parameters are allowed to vary, the KL divergence is relatively small but remains above $10^{-10}$. For the smallest neighborhood ($c=0.1$), the empirical distance is slightly below 0.05 at $T=80$ and crosses 0.05 only as $T$ exceeds 200; for larger neighborhoods, the threshold is crossed at smaller sample sizes. This suggests that the model is difficult to distinguish from neighboring parameterizations at limited sample sizes. Notably, the parameter $\beta$ binds the constraint for neighborhood sizes of 0.1 and 0.5, while $\phi_\pi$ binds it for $c=1$. This difference arises from setting the bound for $\beta$ to $[0.1, 0.999]$ to maintain economic interpretability, which prevents $\beta$ from binding the constraint at $c=1$. The results therefore indicate that the discount factor plays the most significant role in the difficulty of global identification, with $\phi_\pi$, the monetary policy response coefficient to inflation, the next most influential.

Next, I sequentially fix the binding parameters and re-minimize. For $c=0.1$ and $c=0.5$, the binding parameters in sequence are $\beta$, $\phi_\pi$, and $\sigma_a$; for $c=1$, where $\beta$'s bounds prevent it from binding, the sequence is $\phi_\pi$, $\sigma_a$, and $\sigma_g$. After fixing the first binding parameter (column b of Tables~\ref{tab: optimizer_RE} and~\ref{tab: KL_RE}), the empirical distance exceeds 0.05 even at $T=80$ across all neighborhood sizes; at $T=150$, it reaches 0.5810 for $c=0.5$ and 0.7507 for $c=1$. After fixing the second binding parameter (column c), the empirical distance further increases: at $T=150$, it reaches 0.7393 for $c=0.5$ and 0.9989 for $c=1$.

\begin{table}[h!]
\centering
\caption{Parameter values minimizing the KL criterion, HSY (2024) model under RE}
 \label{tab: optimizer_RE}
\vspace*{0.2cm}
\begin{tabular}{c c c c c c c c c c c c}
\toprule
 & \multicolumn{4}{c}{\textbf{(a) All parameters can vary}} & \multicolumn{4}{c}{\textbf{(b) $\beta (\phi_\pi$ for $ c=1 )$ fixed}} & \multicolumn{3}{c}{\textbf{(c)$\beta$\&$\phi_\pi$ ($\phi_\pi$\&$\sigma_a$)fixed}} \\
\cmidrule(lr){2-5} \cmidrule(lr){6-9} \cmidrule(lr){10-12}
 & $\gamma_0^{RE}$ & $c=0.1$ & $c=0.5$ & $c=1$ & $c=0.1$ & $c=0.5$ & $c=1$ & & $c=0.1$ & $c=0.5$ & $c=1$ \\
\midrule
$\phi_y$ & 0.08 & 0.08 & 0.08 & 0.13 & 0.09 & 0.04 & 0.04 & & 0.08 & 0.06 & 0.05 \\
$\phi_{\pi}$ & 1.21 & 1.18 & 1.09 & \textbf{0.21} & \textbf{1.11} & \textbf{1.71} & 1.21 & & 1.21 & 1.21 & 1.21 \\
$\beta$ & 0.99 & \textbf{0.89} & \textbf{0.49} & 0.35 & 0.99 & 0.99 & 0.10 & & 0.99 & 0.99 & 0.999 \\
$\kappa$ & 0.15 & 0.15 & 0.16 & 0.10 & 0.13 & 0.26 & 0.12 & & 0.13 & 0.09 & 0.14 \\
$\rho_a$ & 0.56 & 0.56 & 0.55 & 0.57 & 0.56 & 0.52 & 0.53 & & 0.55 & 0.52 & 0.60 \\
$\rho_g$ & 0.95 & 0.95 & 0.94 & 0.91 & 0.95 & 0.95 & 0.94 & & 0.95 & 0.96 & 0.97 \\
$\sigma_a$ & 0.91 & 0.98 & 1.20 & 1.77 & 0.98 & 0.77 & \textbf{1.91} & & \textbf{1.01} & \textbf{1.41} & 0.91 \\
$\sigma_g$ & 1.54 & 1.48 & 1.30& 1.17 & 1.54 & 1.54 & 1.08 & & 1.54 & 1.57 & \textbf{2.54} \\
$\sigma_m$ & 0.39 & 0.39 & 0.38 & 0.38 & 0.39 & 0.41 & 0.39 & & 0.39 & 0.39 & 0.38 \\
\bottomrule
\end{tabular}
\par
\vspace*{0.3cm}
\begin{minipage}{1.02\textwidth}
\justify
{\small Note: KL denotes $KL_{ff}(\gamma_0^{RE}, \gamma_c^{RE})$ with $\gamma_0^{RE}$  corresponding to the benchmark specification. The values are rounded to the second decimal place except for $\beta$. The bold value signifies the binding constraint. }
\end{minipage}
\end{table}

\begin{table}[h!] 
\centering
\caption{KL and empirical distances between $\gamma_c$ and $\gamma_0$, HSY (2024) model under RE}
\label{tab: KL_RE}
\vspace*{0.2cm}
\setlength{\tabcolsep}{3pt} 
\begin{tabular}{c c c c c c c c c c c c}
\toprule
 & \multicolumn{4}{c}{\textbf{(a)\small All parameters can vary}} & \multicolumn{3}{c}{\textbf{(b) $\beta (\phi_\pi$ for $ c=1 )$ fixed}} & \multicolumn{4}{c}{\textbf{(c) $\beta$\&$\phi_\pi$($\phi_\pi$\&$\sigma_a$)fixed}} \\
\cmidrule(lr){2-5} \cmidrule(lr){6-8} \cmidrule(lr){10-12}
 & $c=0.1$ & $c=0.5$ & $c=1$ & & $c=0.1$ & $c=0.5$ & $c=1$ & & $c=0.1$ &$c=0.5$ & $c=1$  \\
\midrule
KL & 5.06E-05 & 1.45E-03 & 0.0160 & & 3.59E-04 & 1.31E-02 & 0.0202 & & 0.0011 & 0.0155 & 0.1389 \\
$T=80$ & 0.0463 & 0.0501 & 0.2449 & & 0.0824 & 0.3662 & 0.5231 & & 0.1213 & 0.5434 & 0.9827 \\
$T=150$ & 0.0497 & 0.0730 & 0.4744 & & 0.0966 & 0.5810 & 0.7507 & & 0.1551 & 0.7393 & 0.9989 \\
$T=200$ & 0.0516 & 0.0894 & 0.6161 & & 0.1055 & 0.6978 & 0.8489 & & 0.1770 & 0.8280 & 0.9998 \\
$T=1000$ & 0.0730 & 0.3677 & 0.9999 & & 0.2165 & 0.9997 & 0.9999 & & 0.4542 & 0.9999 & 1.0000 \\
\bottomrule
\end{tabular}
\par
\vspace*{0.3cm}
\begin{minipage}{1.02\textwidth}
\justify
{\small Note: KL denotes $KL_{ff}(\gamma_0^{RE}, \gamma_c^{RE})$ with $\gamma_0^{RE}$  given in the columns of Table~\ref{tab: optimizer_RE}. The empirical distance measure equals $p_{ff}(\gamma_0^{RE}, \gamma_c^{RE}, 0.05, T)$, where T is specified in the last four rows of the table. }
\end{minipage}
\end{table}
\clearpage
\newpage

\subsubsection*{B.3 Robustness check: Identification of parameters estimated via MCMC }\label{sec: App_B.3}

This subsection examines whether the identification results are robust to using an alternative benchmark parameter vector estimated by Bayesian MCMC. The priors are the same as those used in the SMC estimation, reported in Table~\ref{tab:prior_posterior}. The MCMC posterior estimates are reported in Table~\ref{tab:posterior_MCMC}.

\begin{table}[h] 
\centering
\caption{Posterior Distribution }
\label{tab:posterior_MCMC}
\vspace*{0.2cm}
\setlength{\tabcolsep}{10pt} 
\begin{tabular}{p{3cm} p{3cm} p{3cm}}
\toprule
\textbf{Parameter} & \textbf{Mean} & \textbf{[05, 95]}  \\
\midrule
$\theta$ & 0.56 & [0.44, 0.70]  \\
$\phi_y$ & 0.10 & [0.07, 0.13] \\
$\phi_{\pi}$ & 1.14 & [1.00, 1.26]  \\
$\kappa$ & 0.13 & [0.08, 0.17]  \\
$\rho_a$ & 0.80 & [0.65, 0.91] \\
$\rho_g$ & 0.94 & [0.92, 0.96]  \\
$\sigma_a$ & 0.55 & [0.39, 0.74] \\
$\sigma_g$ & 1.77 & [1.49, 2.08]  \\
$\sigma_m$ & 0.38 & [0.33, 0.45]  \\
\bottomrule
\end{tabular}
\par
\vspace*{0.3cm}
\begin{minipage}{0.7\textwidth}
\justify
{\small Note: The results were estimated using Dynare version 6.2 with a type of MCMC, slice sampling. The number of replication draws is set to 700, which, according to \cite{planas2015slice}, is approximately equivalent to 50,000 draws using classical Metropolis-Hastings sampling. The number of replication blocks is set to 1.
}
\end{minipage}
\end{table}
Tables~\ref{tab: optimizer_MCMC} and~\ref{tab: KL_MCMC} report the corresponding identification results. The MCMC-based benchmark yields a similar identification profile to the SMC benchmark. When all parameters are allowed to vary, the weakest-identified parameters remain the shock standard deviations. When $\sigma_a$ and $\sigma_g$ are fixed, the diagnosticity parameter $\theta$ becomes the binding parameter in some neighborhoods. Nevertheless, the empirical distance remains large: for $c=0.5$ and $T=80$, it reaches 0.7966, well above the 0.05 threshold. These results confirm that $\theta$ remains globally identified under the MCMC-based benchmark. The small differences between the SMC- and MCMC-based results reflect the different posterior benchmark points at which the identification exercise is evaluated, not a change in the underlying identification conclusion.

\begin{table}[!h]
\centering
\caption{Parameter values minimizing the KL criterion, HSY (2024) model under DE}
\label{tab: optimizer_MCMC}
\vspace*{0.2cm}
\begin{tabular}{c c c c c c c c c c c c}
\toprule
 & \multicolumn{4}{c}{\textbf{(a) All parameters can vary}} & \multicolumn{4}{c}{\textbf{(b) $\sigma_a$ fixed}} & \multicolumn{3}{c}{\textbf{(c) $\sigma_a$ and $\sigma_g$ fixed}} \\
\cmidrule(lr){2-5} \cmidrule(lr){6-9} \cmidrule(lr){10-12}
 & $\gamma_0^{MCMC}$ & $c=0.1$ & $c=0.5$ & $c=1$ & $c=0.1$ & $c=0.5$ & $c=1$ & & $c=0.1$ & $c=0.5$ & $c=1$ \\
\midrule
$\theta$ & 0.56 & 0.61 & 0.76 & 0.84 & 0.60 & 0.47 & 0.42 & & 0.58 & \textbf{0.06} & \textbf{1.56} \\
$\phi_y$ & 0.10 & 0.10 & 0.11 & 0.11 & 0.11 & 0.09 & 0.09 & & 0.11 & 0.09 & 0.18 \\
$\phi_{\pi}$ & 1.14 & 1.06 & 0.75 & 0.38 & 1.13 & 1.12 & 1.10 & & \textbf{1.04} & 1.07 & 0.68 \\
$\beta$ & 0.990 & 0.894 & 0.644 & 0.502 & 0.966 & 0.999 & 0.999 & & 0.974 & 0.999 & 0.72 \\
$\kappa$ & 0.13 & 0.12 & 0.10 & 0.07 & 0.14 & 0.12 & 0.11 & & 0.11 & 0.20 & 0.13 \\
$\rho_a$ & 0.80 & 0.80 & 0.79 & 0.78 & 0.81 & 0.81 & 0.83 & & 0.82 & 0.74 & 0.87 \\
$\rho_g$ & 0.94 & 0.94 & 0.93 & 0.92 & 0.94 & 0.95 & 0.95 & & 0.94 & 0.95 & 0.83 \\
$\sigma_a$ & 0.55 & \textbf{0.65} & \textbf{1.05} & \textbf{1.55} & 0.55 & 0.55 & 0.55 & & 0.55 & 0.55 & 0.55 \\
$\sigma_g$ & 1.77 & 1.70 & 1.54 & 1.48 & \textbf{1.67} & \textbf{2.27} & \textbf{2.77} & & 1.77 & 1.77 & 1.77 \\
$\sigma_m$ & 0.38 & 0.38 & 0.37 & 0.37 & 0.38 & 0.38 & 0.38 & & 0.38 & 0.37 & 0.39 \\
\bottomrule
\end{tabular}
\par
\vspace*{0.3cm}
\begin{minipage}{1.02\textwidth}
\justify
{\small Note: KL denotes $KL_{ff}(\gamma_0^{MCMC}, \gamma_c^{MCMC})$ with $\gamma_0^{MCMC}$  corresponding to the benchmark speciﬁcation. The values are rounded to the second decimal place except for $\beta$. The bold value signiﬁes the binding constraint. }
\end{minipage}
\end{table}

\begin{table}[h!]
\centering
\caption{KL and empirical distances between $\gamma_c$ and $\gamma_0$, HSY (2024) model}
\label{tab: KL_MCMC}
\vspace*{0.2cm}
\setlength{\tabcolsep}{3pt} 
\begin{tabular}{c c c c c c c c c c c c}
\toprule
 & \multicolumn{4}{c}{\textbf{(a)\small All parameters can vary}} & \multicolumn{3}{c}{\textbf{(b) $\sigma_a$ fixed}} & \multicolumn{3}{c}{\textbf{(c) $\sigma_a$ and $\sigma_g$ fixed}} \\
\cmidrule(lr){2-5} \cmidrule(lr){6-8} \cmidrule(lr){10-12}
 & $c=0.1$ & $c=0.5$ & $c=1$ & & $c=0.1$ & $c=0.5$ & $c=1$ & & $c=0.1$ & $c=0.5$ & $c=1$  \\
\midrule
KL & 1.15E-04 & 2.51E-03 & 6.94E-03 & & 9.19E-04 & 0.0224 & 0.0718 & & 2.09E-03 & 0.0385 & 0.1541 \\
$T=80$ & 0.0535 & 0.0964 & 0.1912 & & 0.0856 & 0.7032 & 0.9664 & & 0.1689 & 0.7966 & 0.9982 \\
$T=150$ & 0.0592 & 0.1451 & 0.3184 & & 0.1106 & 0.8739& 0.9975 & & 0.2262 & 0.9552 & 0.9999 \\
$T=200$ & 0.0627 & 0.1794 & 0.4044 & & 0.1270 & 0.9327 & 0.9996 & & 0.2635 & 0.9860 & 1.0000 \\
$T=1000$ & 0.1032 & 0.6467 & 0.9701 & & 0.3536 & 1.0000 & 1.0000 & & 0.6784 & 1.0000 & 1.0000 \\
\bottomrule
\end{tabular}
\par
\vspace*{0.3cm}
\begin{minipage}{1.02\textwidth}
\justify
{\small Note: KL denotes $KL_{ff}(\gamma_0^{MCMC}, \gamma_c^{MCMC})$ with $\gamma_0^{MCMC}$  given in the columns of Table~\ref{tab: KL_MCMC}. The empirical distance measure equals $p_{ff}(\gamma_0^{MCMC}, \gamma_c^{MCMC}, 0.05, T)$, where T is specified in the last four rows of the table. }
\end{minipage}
\end{table}

\section{A medium scale DSGE}
\subsection{The model} \label{sec: App_C.1}
In this section I list the equations I used in section \ref{Identification_med_FD} which are same as the equations in the replication Dynare code of \cite{l2024incorporating} except the information friction part.\\
\textbf{Nonstationary Productivity}:\\
Productivity (in logs) is given by the sum of two components:
\begin{equation*}
a_t = x_t + z_t.
\end{equation*}
The permanent component, $x_t$, follows a unit root process given by
\begin{equation*}
\Delta x_t = \rho_x \Delta x_{t-1} + \varepsilon_{x,t}.
\end{equation*}
The transitory component, $z_t$, follows a stationary process given by
\begin{equation*}
z_t = \rho_z z_{t-1} + \varepsilon_{z,t}.
\end{equation*}
\cite{blanchard2013news} assume $a_t$ is a unit root process 
\begin{equation}\label{equ: a_t}
a_t = a_{t-1} + \varepsilon_{a,t},
\end{equation}
with the variance of $\varepsilon_{a,t}$ equal to $\sigma^2_a$. In general, a given univariate process is consistent with an infinity of decompositions between a permanent and a transitory component with orthogonal innovations. \cite{blanchard2013news} choose one-parameter family which deliver the above univariate random walk:
\begin{equation*}
\rho_x = \rho_z = \rho, \qquad \sigma^2_x = (1-\rho)^2 \sigma^2_a, \qquad \sigma^2_z = \rho\sigma^2_a,
\end{equation*}
Consumers observe current and past productivity, $a_t$. In addition, I assume they receive a signal about permanent productivity growth.\footnote{\cite{l2024incorporating} assume that consumers observe a signal on the level of the permanent productivity component. I modify this assumption to stabilize the general equilibrium system; otherwise, the optimization converges very slowly due to the presence of a unit root.}
\begin{equation*}
s_t = \Delta x_t + \varepsilon_{s,t},
\end{equation*}
where $\varepsilon_{s,t}$ is i.i.d. normal with variance $\sigma^2_s$. Moreover, consumers know the structure of the model, i.e., know $\rho$ and the variances of the three shocks.\\
\textbf{Kalman Filter:}

Following \cite{l2024incorporating}, I employ the Kalman filter as a computational tool to transform an incomplete information model into a form that mimics complete information while preserving the economic intuition of agents learning from signals. Since the global identification condition requires the spectrum to be nonsingular, I follow \cite{qu2023using} and define the unit root variable in growth rates. The state equations are:

\begin{align*}
   &\Delta x_t = \rho_x \Delta x_{t-1} + \varepsilon_{x,t}\\
   &z_t = \rho_z z_{t-1} + \varepsilon_{z,t}\\
   &z_{t-1} = z_{t-1}.
\end{align*}

The observation equations are:

\begin{align*}
   &\Delta a_t = \Delta x_t + z_t - z_{t-1}\\
   & s_t = \Delta x_t + \varepsilon_{s,t}.
\end{align*}

I write the above system in matrix form as:

\begin{equation}
     \mathbf{X}_t = \mathbf{F} \mathbf{X}_{t-1} + \boldsymbol{\varepsilon}_t
\end{equation}

where

\begin{equation*}
\mathbf{X}_t = \begin{bmatrix}
\Delta x_t \\
z_t \\
z_{t-1}
\end{bmatrix}, \quad
\mathbf{F} = \begin{bmatrix}
\rho_x & 0 & 0 \\
0 & \rho_z & 0 \\
0 & 1 & 0
\end{bmatrix}, \quad
\boldsymbol{\varepsilon}_t = \begin{bmatrix}
\varepsilon_{x,t} \\
\varepsilon_{z,t}\\
0
\end{bmatrix},
\end{equation*}

and

\begin{equation}
    \mathbf{Y}_t = \mathbf{A} \mathbf{X}_t + \boldsymbol{\eta}_t
\end{equation}

where

\begin{equation*}
\mathbf{Y}_t = \begin{bmatrix}
\Delta a_t \\
 s_t
\end{bmatrix}, \quad
\mathbf{A} = \begin{bmatrix}
1 & 1 & -1 \\
1 & 0 & 0
\end{bmatrix}, \quad
\boldsymbol{\eta}_t = \begin{bmatrix}
0 \\
\varepsilon_{s,t}
\end{bmatrix}.   
\end{equation*}

Following \cite{l2024incorporating}, I employ the Kalman filter as a computational tool to transform an incomplete information model into a form that mimics complete information:
\begin{align}\label{eq:state update}
\begin{pmatrix}
\Delta x_{t|t} \\
z_{t|t} \\
z_{t-1|t}
\end{pmatrix}
&= (\boldsymbol{I} - \boldsymbol{K A}) \boldsymbol{F }
\begin{pmatrix}
\Delta x_{t-1|t-1} \\
z_{t-1|t-1} \\
z_{t-2|t-1}
\end{pmatrix}
+ \boldsymbol{K}
\begin{pmatrix}
\Delta a_t \\
s_t
\end{pmatrix} \\ \nonumber
    &= \boldsymbol{F}
\begin{pmatrix}
\Delta x_{t-1|t-1} \\
z_{t-1|t-1} \\
z_{t-2|t-1}
\end{pmatrix}
+ \boldsymbol{ K}\underbrace{ \left[
\begin{pmatrix}
\Delta a_t \\
s_t
\end{pmatrix}
- \boldsymbol{AF}
\begin{pmatrix}
\Delta x_{t-1|t-1} \\
z_{t-1|t-1} \\
z_{t-2|t-1}
\end{pmatrix}
\right]}_{\equiv \boldsymbol{e_t}},
\end{align}

\begin{align}\label{eq:observable update}
\Rightarrow \quad 
\begin{pmatrix}
\Delta a_t \\
s_t
\end{pmatrix}
= \boldsymbol{AF}
\begin{pmatrix}
\Delta x_{t-1|t-1} \\
z_{t-1|t-1} \\
z_{t-2|t-1}
\end{pmatrix}
+ \boldsymbol{e_t},
\end{align}
where $\mathbf{K}$ the steady-state, time-invariant Kalman gain obtained from the Riccati recursion. Equation (\ref{eq:state update}) and (\ref{eq:observable update}) are incorporated into the general equilibrium system.

\begin{align*}
&\hat{\lambda}_t - \hat{G}_{a,t} - \hat{\pi}_t = \hat{i}_t + \mathbb{E}^{\theta}_t[\hat{\lambda}_{t+1} - \hat{G}_{a,t} - \hat{G}_{a,t+1} - \hat{\pi}_t - \hat{\pi}_{t+1}] \\
&\hat{\lambda}_t + \frac{G_a}{G_a - h}\hat{c}_t - \frac{h}{G_a - h}(\hat{c}_{t-1} - \hat{G}_{a,t}) = 0 \\
&\hat{\pi}_t = \beta\mathbb{E}^{\theta}_t[\hat{\pi}_{t+1} - \iota_p\hat{\pi}_t] + \iota_p\hat{\pi}_{t-1} + \frac{\epsilon_p - 1}{\psi_p}\hat{mc}_t + \hat{\lambda}^{p,*}_t \\
&\hat{\pi}^w_t = \beta\mathbb{E}^{\theta}_t[\hat{\pi}^w_{t+1} - \iota_w\hat{\pi}_t - \iota_w\hat{G}_{a,t+1}] + \iota_w\hat{\pi}_{t-1} + \iota_w\hat{G}_{a,t} + \frac{\epsilon_w\omega L^{1+\nu}}{\psi_w}[\nu\hat{L}_t - \hat{w}_t - \hat{\lambda}_t] + \hat{\lambda}^{w,*}_t \\
&\hat{k}^u_{t+1} = \frac{\mathbb{I}}{k^u}(\hat{I}_t + \hat{\mu}_t) + \frac{1-\delta_k}{G_a}(\hat{k}^u_t - \hat{G}_{a,t}) \\
&\hat{q}_t - \hat{G}_{a,t} + \hat{\lambda}_t = \mathbb{E}^{\theta}_t[\hat{\lambda}_{t+1} - \hat{G}_{a,t} - \hat{G}_{a,t+1} + \frac{r^K}{r^K + 1 - \delta_k}\hat{r}^K_{t+1} + \frac{1-\delta_k}{r^K + 1 - \delta_k}\hat{q}_{t+1}] \\
&\hat{q}_t + \hat{\mu}_t - S''(1)(\hat{I}_t - \hat{I}_{t-1} + \hat{G}_{a,t}) + \beta S''(1)\mathbb{E}^{\theta}_t[\hat{I}_{t+1} - \hat{I}_t + \hat{G}_{a,t+1}] = 0 \\
&\hat{k}_t = \hat{u}_t + \hat{k}^u_t - \hat{G}_{a,t} \\
&\hat{r}^K_t = \frac{\chi''(1)}{\chi'(1)}\hat{u}_t \\
&\hat{y}_t = \alpha\hat{k}_t + (1-\alpha)\hat{L}_t \\
&\hat{r}^K_t = \hat{w}_t + \hat{L}_t - \hat{k}_t \\
&\hat{mc}_t = \alpha\hat{r}^K_t + (1-\alpha)\hat{w}_t \\
&\hat{i}_t = \rho_R\hat{i}_{t-1} + (1-\rho_R)(\phi_\pi\hat{\pi}_t + \phi_y\hat{y}_t) + \hat{\lambda}^{mp}_t \\
&\frac{1}{\lambda^g}\hat{y}_t = \frac{c}{y}\hat{c}_t + \frac{\mathbb{I}}{y}\hat{I}_t + \frac{\chi'(1)k}{y}\hat{u}_t + \frac{1}{\lambda^g}\hat{\lambda}^g_t \\
&\hat{\mu}_t = \rho_\mu\hat{\mu}_{t-1} + \varepsilon_{\mu,t} \\
&\hat{\lambda}^{mp}_t = \rho_{mp}\hat{\lambda}^{mp}_{t-1} + \varepsilon_{mp,t} \\
&\hat{\lambda}^g_t = \rho_g\hat{\lambda}^g_{t-1} + \varepsilon_{g,t} \\
&\hat{\lambda}^{p,*}_t = \rho_p\hat{\lambda}^{p,*}_{t-1} + \varepsilon_{p,t} - \phi_p\varepsilon_{p,t-1} \\
&\hat{\lambda}^{w,*}_t = \rho_w\hat{\lambda}^{w,*}_{t-1} + \varepsilon_{w,t} - \phi_w\varepsilon_{w,t-1} 
\end{align*}

\textbf{Disturbances}:

\begin{alignat*}{2}
&\text{TFP growth shock:}\quad && \varepsilon_{x,t} \sim N(0,\sigma^2_x) \\
&\text{Stationary TFP shock:}\quad && \varepsilon_{z,t} \sim N(0,\sigma^2_z) \\
&\text{Noise shock:}\quad && \varepsilon_{s,t} \sim N(0,\sigma^2_s) \\
&\text{MEI shock:}\quad && \varepsilon_{\mu,t} \sim N(0,\sigma^2_\mu) \\
&\text{Monetary policy shock:}\quad && \varepsilon_{mp,t} \sim N(0,\sigma^2_{mp}) \\
&\text{Government spending shock:}\quad && \varepsilon_{g,t} \sim N(0,\sigma^2_g) \\
&\text{Price markup shock:}\quad && \varepsilon_{p,t} \sim N(0,\sigma^2_p) \\
&\text{Wage markup shock:}\quad && \varepsilon_{w,t} \sim N(0,\sigma^2_w) 
\end{alignat*}

\subsection{Bayesian estimation} \label{sec: App_C.2}
\begin{table}[!htbp]
\caption{Posterior distributions of the medium-scale model parameters}
\label{table: posterior_med}
\begin{center}
\begin{tabular}{llcc}
\hline
Parameter & Description & Post. Mean & 90\% HPD Interval \\
\hline
$\theta$ & diagnosticity & 0.72 & [0.58, 0.86] \\
$\alpha$ & cap. share & 0.13 & [0.12, 0.14] \\
$h$ & habits & 0.72 & [0.70, 0.75] \\
$\frac{\chi''(1)}{\chi'(1)}$ & cap. util. costs & 5.09 & [3.62, 6.55] \\
${\psi}_p$ & Rotemberg prices & 122.47 & [95.30, 148.26] \\
${\psi}_w$ & Rotemberg wages & 507.44 & [254.73, 773.38] \\
$\nu$ & inv. Frisch elas. & 3.71 & [2.34, 5.05] \\
$S''(1)$ & inv. adj. costs & 6.93 & [5.93, 7.99] \\
$\rho_R$ & m.p. rule & 0.58 & [0.54, 0.62] \\
$\phi_\pi$ & m.p. rule & 1.54 & [1.42, 1.66] \\
$\phi_x$ & m.p. rule & 0.006 & [0.00, 0.01] \\
\multicolumn{4}{l}{\textit{Technology Shocks}} \\
$\rho$ & persist. & 0.85 & [0.83, 0.87] \\
$\sigma_a$ & tech. shock s.d. & 1.43 & [1.31, 1.55] \\
$\sigma_s$ & noise shock s.d. & 0.29 & [0.23, 0.35] \\
\multicolumn{4}{l}{\textit{Investment-Specific Shocks}} \\
$\rho_\mu$ & persist. & 0.31 & [0.25, 0.35] \\
$\sigma_\mu$ & s.d. & 18.63 & [15.99, 21.82] \\
\multicolumn{4}{l}{\textit{Mark-up Shocks}} \\
$\rho_p$ & persist. & 0.88 & [0.83, 0.92] \\
$\phi_p$ & ma. comp. & 0.58 & [0.46, 0.70] \\
$\sigma_p$ & s.d. & 0.16 & [0.13, 0.19] \\
$\rho_w$ & persist. & 0.997 & [0.99, 1.00] \\
$\phi_w$ & ma. comp. & 0.54 & [0.39, 0.66] \\
$\sigma_w$ & s.d. & 0.44 & [0.35, 0.53] \\
\multicolumn{4}{l}{\textit{Policy Shocks}} \\
$\rho_{mp}$ & persist. & 0.03 & [0.01, 0.05] \\
$\sigma_{mp}$ & s.d. & 0.38 & [0.34, 0.42] \\
$\rho_g$ & persist. & 0.94 & [0.91, 0.96] \\
$\sigma_g$ & s.d. & 0.37 & [0.34, 0.40] \\
\multicolumn{4}{l}{\textit{Measurement Errors}} \\
$\sigma_{ygr}$ & s.d. & 0.50 & [0.45, 0.55] \\
$\sigma_{cgr}$ &  s.d. & 0.41 & [0.36, 0.46] \\
$\sigma_{igr}$ &  s.d. & 1.44 & [1.26, 1.61] \\
$\sigma_{\pi}$ &  s.d. & 0.27 & [0.24, 0.30] \\
$\sigma_{\hat{i}}$ & s.d. & 0.16 & [0.14, 0.18] \\
\hline
\end{tabular}
\par
\vspace*{0.3cm}
\begin{minipage}{0.95\textwidth}
\justify
{\small Note: This table shows the posterior distribution under DE. The values are rounded to two decimal places except for $\phi_{x}$ and $\rho_w$.}
\end{minipage}
\end{center}
\end{table}

\clearpage
\newpage

\subsection{Identification of frictions in the RE model}
\label{sec: App_C.3}
\begin{table}[h!]
\centering
\caption{Closest RE counterparts to the RE benchmark with constrained frictions}
\begin{tabular}{lccccccc}
\hline
  & $\gamma_0^{med,RE}$ & $\sigma_s=0.1$ & $\kappa_p=1$ & $\kappa_w=1$ & $h=0.1$ & $S''=1$ & $\frac{\chi''(1)}{\chi'(1)}=1$ \\
\hline
KL & -- & 0.0084 & 0.9659 & 0.5446 & 0.5990 & 1.2369 & 1.1633 \\
$T=80$ & -- & 0.3296 & 1.0000 & 0.9997 & 1.0000 & 1.0000 & 0.6266 \\
$T=150$ & -- & 0.4896 & 1.0000 & 1.0000 & 1.0000 & 1.0000 & 0.6709 \\
$\alpha$ & 0.14 & 0.14 & 0.14 & 0.13 & 0.12 & 0.16 & 0.20 \\
$h$ & 0.58 & 0.57 & 0.60 & 0.48 & -- & 0.52 & 0.50 \\
$\frac{\chi''(1)}{\chi'(1)}$ & 5.55 & 5.28 & 4.84 & 4.46 & 5.34 & 5.83 & -- \\
$\kappa_p$ & 0.03 & 0.03 & -- & 0.05 & 0.05 & 0.03 & 0.07 \\
$\kappa_w$ & 0.001 & 0.002 & 0.001 & -- & 0.12 & 0.002 & 0.17 \\
$\nu$ & 1.28 & 0.80 & 2.00 & 0.50 & 1.38 & 0.50 & 0.50 \\
$S''$ & 7.00 & 6.81 & 6.53 & 6.67 & 6.52 & -- & 6.79 \\
$\rho_R$ & 0.68 & 0.68 & 0.68 & 0.57 & 0.44 & 0.64 & 0.10 \\
$\phi_\pi$ & 1.04 & 1.12 & 1.02 & 1.24 & 1.22 & 1.01 & 1.05 \\
$\phi_x$ & 0.001 & 0.001 & 0.001 & 0.0001 & 0.0002 & 0.011 & 0.0001 \\
$\rho$ & 0.96 & 0.96 & 0.94 & 0.92 & 0.89 & 0.96 & 0.94 \\
$\rho_\mu$ & 0.34 & 0.34 & 0.32 & 0.21 & 0.29 & 0.10 & 0.37 \\
$\rho_p$ & 0.81 & 0.81 & 0.90 & 0.87 & 0.89 & 0.78 & 1.00 \\
$\phi_p$ & 0.54 & 0.55 & 0.10 & 0.60 & 0.67 & 0.52 & 0.10 \\
$\rho_w$ & 0.68 & 0.67 & 0.65 & 0.99 & 0.99 & 0.57 & 0.99 \\
$\phi_w$ & 0.57 & 0.56 & 0.89 & 0.10 & 0.39 & 0.36 & 0.36 \\
$\rho_{mp}$ & 0.02 & 0.01 & 0.04 & 0.01 & 0.01 & 0.01 & 0.69 \\
$\rho_g$ & 0.90 & 0.90 & 0.90 & 0.90 & 0.92 & 0.90 & 0.99 \\
$\sigma_a$ & 1.59 & 1.59 & 1.58 & 1.61 & 1.43 & 1.61 & 1.02 \\
$\sigma_s$ & 0.33 & -- & 0.32 & 0.10 & 0.10 & 0.18 & 0.21 \\
$\sigma_\mu$ & 19.70 & 19.24 & 18.61 & 21.57 & 19.47 & 15.00 & 18.71 \\
$\sigma_p$ & 0.20 & 0.20 & 1.38 & 0.20 & 0.20 & 0.20 & 0.30 \\
$\sigma_w$ & 0.46 & 0.46 & 0.72 & 3.00 & 0.78 & 0.49 & 0.79 \\
$\sigma_{mp}$ & 0.33 & 0.33 & 0.33 & 0.35 & 0.39 & 0.33 & 0.44 \\
$\sigma_g$ & 0.37 & 0.37 & 0.37 & 0.37 & 0.37 & 0.37 & 0.46 \\
\hline
\end{tabular}
\label{tab: friction_comparison_RERE}
\par
\vspace*{0.3cm}
\begin{minipage}{\textwidth}
\justify
{\small Note: KL (row 2) and the empirical distance measures (rows 3-4) are defined as 
$KL_{ff}(\gamma_0^{med,RE}, \gamma^{med,RE})$ and $p_{ff}(\gamma_0^{med,RE}, \gamma^{med,RE}, 0.05, T)$, 
where the criterion is computed over the full frequency range. Each column reports the parameter 
vector $\gamma^{med,RE}$ that minimizes the KL divergence from the RE benchmark parameter vector 
$\gamma_0^{med,RE}$ (column 1), subject to the friction restriction indicated in the column header. 
The diagnostic expectation parameter $\theta$ is fixed at 0 throughout. All parameter values 
are rounded to two decimal places, except $\phi_x$.}
\end{minipage}
\end{table}


\end{document}